\documentclass[prl,showpacs]{revtex4}
\usepackage{graphics}
\usepackage{epsfig}
\usepackage{graphicx}

\usepackage{amsfonts} 

\newcommand{\ba}{\begin{eqnarray}}
\newcommand{\ea}{\end{eqnarray}}
\newcommand{\ua}{\uparrow}
\newcommand{\da}{\downarrow}

\begin{document}

\title{Macroscopic Distinguishability Between Quantum States Defining Different Phases of Matter: Fidelity and the Uhlmann Geometric Phase}

\author{Nikola Paunkovi\'c}
\affiliation{SQIG -- Instituto de Telecomunica\c{c}\~oes, IST, Lisbon, P-1049-001 Lisbon, Portugal}
\author{Vitor Rocha Vieira}
\affiliation{CFIF and Department of Physics, IST, Technical University of Lisbon, P-1049-001 Lisbon, Portugal}

\date{\today}

\begin{abstract}
We study the fidelity approach to quantum phase transitions (QPTs)
and apply it to general thermal phase transitions (PTs). We analyze
two particular cases: the Stoner-Hubbard  itinerant electron
model of magnetism and the BCS theory of superconductivity. In both cases we
show that the sudden drop of the mixed state fidelity marks the line
of the phase transition. We conduct a detailed analysis of the
general case of systems given by mutually commuting Hamiltonians,
where the non-analyticity of the fidelity is directly related to the
non-analyticity of the relevant response functions (susceptibility
and heat capacity), for the case of symmetry-breaking transitions.
Further, on the case of BCS theory of superconductivity, given by mutually
non-commuting Hamiltonians, we analyze the structure of the system's
eigenvectors in the vicinity of the line of the phase transition
showing that their sudden change is quantified by the emergence of a generically
non-trivial Uhlmann mixed state geometric phase.

\end{abstract}

\maketitle

One of the main characteristics of quantum mechanics that makes it
different from any classical physical theory is that in quantum
mechanics two different quantum states, being either pure or mixed,
are in general not fully distinguishable. By fully distinguishable
we mean that it is possible, upon a result of a \emph{single-shot}
measurement of a suitable observable, to infer with probability one
in which of the two given quantum states the observed system has
been prepared. In particular, two pure quantum states are fully
distinguishable if and only if they are orthogonal to each other.
Otherwise, the maximal probability to unambiguously distinguish
between two non-orthogonal pure quantum states is always strictly
smaller than one. The reason for this lies in the fact that, while
the outcomes of measurements on classical systems are, at least in
principle, given with certainty, quantum measurements in general
generate non-trivial probability distributions. This feature of
quantum mechanics has found numerous applications within the field of
quantum information and computation, in particular in quantum
cryptography, quantum communication complexity, designing novel
quantum algorithms, etc. (for an overview, see \cite{nielsen}).

Within the field of quantum information, the function widely used to
quantify the distinguishability between two quantum states
$\hat{\rho}_1$ and $\hat{\rho}_2$ is {\em fidelity} \cite{wootters}, given by the
expression:
\begin{equation}
\label{fidelity-def}
F(\hat{\rho}_1, \hat{\rho}_2) = \mbox{Tr} \sqrt{\sqrt{\hat{\rho}_1}\hat{\rho}_2\sqrt{\hat{\rho}_1}}.
\end{equation}
Note that in the case of pure states $\hat{\rho}_1 =
|\psi_1\rangle\langle \psi_1|$ and $\hat{\rho}_2 =
|\psi_2\rangle\langle \psi_2|$, the above expression reduces to
$F(|\psi_1\rangle\langle\psi_1|, |\psi_2\rangle\langle\psi_2|) =
|\langle\psi_1| \psi_2\rangle|$, which is nothing but the square
root of the probability for a system in state $|\psi_2\rangle$ to
pass the test of being in state $|\psi_1\rangle$. The fidelity
(\ref{fidelity-def}) between two quantum states, given for two
systems $1$ and $2$, quantifies the statistical distinguishability
between them, in a sense of \textit{classical} statistical
distinguishability between the probability distributions obtained by
measuring an \textit{optimal} observable in states $\hat{\rho}_1$
and $\hat{\rho}_2$. In other words, for every observable $\hat{A}$,
we have $F(\hat{\rho}_1, \hat{\rho}_2)\leq
F_c(\{p_1(i|\hat{A})\},\{p_2(i|\hat{A})\})\equiv \sum_i
\sqrt{p_1(i|\hat{A})p_2(i|\hat{A})}$, where
$\{p_{\alpha}(i|\hat{A})\}$, $\alpha\in\{1,2\}$, is a probability
distribution obtained measuring the observable $\hat{A}$ in the
state $\hat{\rho}_{\alpha}$, and
$F_c(\{p_1(i|\hat{A})\},\{p_2(i|\hat{A})\})$ is the classical
fidelity between the two probability distributions $\{p_1(i|\hat{A})\}$
and $\{p_2(i|\hat{A})\}$. For an overview of the results on
distinguishability between quantum states and its applications to
the field of quantum information, see \cite{fuchs} and the
references therein.

Quantum mechanics was originally developed to
describe the behavior of microscopic systems. Therefore, most of
its applications are focused on the study of the properties and
dynamics of quantum states referring to such systems, where quantum
features dominate. On the other hand, it is a
common assumption that classical behavior emerges in the
thermodynamical limit, when the number of degrees of
freedom of the system becomes large (in the limit of a large number of microscopic
sub-systems). Yet, there is no special objection why quantum
mechanics should not be generally applicable, even to macroscopic
systems. Indeed, macroscopic phenomena such as magnetism,
superconductivity or superfluidity, to name just a few, can only be
explained by using the rules of quantum mechanics. As in these, as
well as in many other, highly physical relevant cases, the
macroscopic features of matter are given through the features of its
quantum states, the question of quantifying those macroscopic
properties given by many-body quantum states arises as a relevant
problem in physics.

In this study, we are interested in those macroscopic features of
matter that define its thermodynamical phase. Different phases of matter are
separated by the so-called {\em regions of criticality}, regions in
parameter space where the system's free energy becomes non-analytic. As
the free energy is a function of the system's states, it is precisely
the features of its states that determine the phase the system is
in. Indeed, different phases have different values of the {\em order
parameter}, given by the expectation value of a certain
observable. As in the case of general
quantum states, here as well, fidelity can be used as a function
whose behavior can mark the regions of criticality (and therefore
the phase transitions).

In the case of quantum phase transitions (QPTs) \cite{sachdev}, which occur
at zero temperature and are driven by purely quantum fluctuations,
the study of the ground state fidelity has been first conducted on the
examples of the Dicke and $XY$ models \cite{us}. Note that in this case,
the ground states are pure quantum states, so the fidelity is given by a
simple overlap between two pure states. It was shown that approaching the regions of criticality the fidelity
between two neighboring ground states exhibits a dramatic drop
 \footnote{Fidelity is closely related to various distance measures on the set of
quantum states: the closer two states are, the more similar (less
distinguishable) they are (for an overview of different distance
measures induced by the fidelity, see \cite{fuchs}).}. Subsequently,
the fidelity approach to QPTs was applied to free Fermi systems and
graphs \cite{zanardi-free_fermion} and to the Bose-Hubbard model
\cite{buonsante-prl}. The connection between fidelity, scaling behavior in QPTs and the renormalization group flows was introduced in \cite{zhou-1} and further discussed in \cite{zhou-2}. Also, it was shown that the fidelity can mark the
regions of criticality in systems whose QPTs can not be described in
terms of Landau-Ginzburg-Wilson (LGW) theory: when the ground states
are given by the matrix product states \cite{zanardi-matrix_prod},
in the case of a topologically ordered QPT
\cite{hamma-topological_order}, and in the Kosterlitz-Thouless type
of transition \cite{gu-kostrelitz} as well. The formal
differential-geometry description of the fidelity approach to QPTs
was first introduced in \cite{zanardi-differential} and subsequently
developed in \cite{zanardi-scaling}, where the connection to the
Berry phase approach to QPTs (see \cite{qpt-berry}) was also
established. Further, on the example of the spin one-half $XXZ$ Heisenberg chain, it was shown that the fidelity does not necessarily exhibit a dramatic drop at the critical point, but that the proper finite-size scaling analysis allows for correct identification of the QPT, see \cite{zanardi-scaling}.
An interesting example of a Heisenberg chain where the
fidelity approach fails when applied to the ground states, but does
mark the point of criticality when applied to the first excited
states, was discussed in \cite{chen-excited} (note that the numerical results were obtained for up to $12$ spins only, which leaves open the question of the ground state fidelity behaviour in the thermodynamic limit).
Introducing the temperature as an
additional parameter, QPTs were studied in \cite{zanardi-thermal}
and it was shown that extending the fidelity approach to general
mixed (thermal) states can still mark the regions of criticality as
well as the cross-over regions at finite temperatures. Finally, the
genuine thermal PTs were discussed in \cite{zanardi-differential}
and \cite{wen-long-thermal}, where the connection between the
singularities in fidelity and specific heat or magnetic
susceptibility was explicitly shown for the cases of systems given
by mutually commuting Hamiltonians and symmetry-breaking PTs of LGW
type.

In this paper, we apply the fidelity approach to general thermal
phase transitions \footnote{The use of the fidelity approach to thermal phase transitions was also suggested in \cite{zhou-1}.}. We analyze in detail two particular examples given by the
Stoner-Hubbard model for magnetism and the BCS theory of superconductivity.
In general, a system is defined by a Hamiltonian $\hat{H}(U)$ which
is a function of a set of parameters representing the interaction
coupling constants generically denoted as $U$. In thermal
equilibrium, a system's state is given by a density operator
$\hat{\rho}(T,U)$. Thus, in discussing the general, thermal as well
as quantum phase transitions, we can consider the coupling
constant(s) $U$ and the temperature $T$ to form a ``generalized''
parameter $q=(T,U)$. We consider the behavior of the fidelity
$F(\hat{\rho}(q),\hat{\rho}(\tilde{q}))$ between two equilibrium
thermal states $\hat{\rho}(q)$ and $\hat{\rho}(\tilde{q})$ defined
by two close parameter points $q=q(T,U)$ and $\tilde{q}=q+\delta
q=(T+\delta T,U+\delta U)$. We show that in both models considered,
the PTs are marked by the sudden drop of fidelity in the vicinity of
regions of criticality - a signature of enhanced distinguishability
between two quantum states defining two different phases of matter,
based on both short-range microscopic as well as long-range
macroscopic features. For the general case of mutually commuting
Hamiltonians, we analytically prove in detail that the same holds
for PTs which fall within the symmetry-breaking paradigm described
by the LGW theory (see also \cite{zanardi-differential} and
\cite{wen-long-thermal}). Further, for the case of mutually
non-commuting Hamiltonians, on the example of BCS theory of superconductivity
we show that the non-analyticity of the fidelity is accompanied by
the emergence of a generically non-trivial Uhlmann geometric phase \cite{uhlmann}, the mixed-state generalization
of the Berry geometric phase (for the relation between QPTs and
Berry phases, see \cite{qpt-berry}, \cite{zanardi-differential} and
\cite{zanardi-scaling}).


\section{Stoner-Hubbard itinerant electron model for magnetism}

First, we discuss the case of the Stoner-Hubbard model for itinerant electrons on a lattice given by the Hamiltonian \cite{stoner}:
\begin{equation}
\label{SH_Hamiltonian-original}
   \hat{H}_{SH} = \sum_k \varepsilon_k\left( \hat{c}^\dag_{k\ua}\hat{c}_{k\ua} + \hat{c}^\dag_{k\da}\hat{c}_{k\da}\right) + U\sum_{l}\hat{c}^\dag_{l\ua}\hat{c}_{l\ua}\hat{c}^\dag_{l\da}\hat{c}_{l\da}.
\end{equation}
The anti-commuting fermionic operators $\hat{c}^\dag_{k\sigma}$
represent the free-electron momentum Bloch modes
($\sigma\in\{\ua,\da\}$), while the on-site operators
$\hat{c}^\dag_{l\sigma}=V^{-1/2}\sum_{k}e^{-ikx_l}\hat{c}^\dag_{k\sigma}$
are given by their Fourier transforms, where $x_l$ represents the
position of the $l$-th lattice site. The coupling constant $U>0$
defines the on-site electron Coulomb repulsion. There are in total
$N$ electrons and they occupy the volume $V$ (such that in the
thermodynamic limit, when $N,V\rightarrow\infty$, we have
$N/V\rightarrow \mbox{const}$). Finally, we assume for simplicity
that the kinetic energy is given by $\varepsilon_k=\hbar^2k^2/(2m)$,
while the number operators are
$\hat{n}_{k\sigma}=\hat{c}^\dag_{k\sigma}\hat{c}_{k\sigma}$ and
$\hat{n}_{l\sigma}=\hat{c}^\dag_{l\sigma}\hat{c}_{l\sigma}=V^{-1}\sum_{qk}e^{iqx_l}\hat{c}^\dag_{k\sigma}\hat{c}_{k+q\sigma}$.
Thus,
\begin{equation}
    \hat{H}_{SH} = \sum_{k\sigma} \varepsilon_k\hat{n}_{k\sigma} + U\sum_l \hat{n}_{l\ua}\hat{n}_{l\da}.
\end{equation}

In order to obtain the mean-field effective Hamiltonian, we neglect
the term quadratic in the fluctuations $\delta\hat{n}_{l\sigma} =
\hat{n}_{l\sigma}-n_{l\sigma}$, where
$n_{l\sigma}=\langle\hat{n}_{l\sigma}\rangle$, in the potential
written as
$\hat{n}_{l\ua}\hat{n}_{l\da}=\delta\hat{n}_{l\ua}\delta\hat{n}_{l\da}+n_{l\ua}\hat{n}_{l\da}+
n_{l\da}\hat{n}_{l\ua}-n_{l\ua}n_{l\da}$.
Expressing the number operators in terms of the Bloch momentum operators,
and using
$\langle\hat{c}^\dag_{k\sigma}\hat{c}_{k^\prime\sigma}\rangle =
\delta_{kk^\prime}\langle\hat{n}_{k\sigma}\rangle$ (which expresses
the translational invariance in a ferromagnetic ground state), the mean-field linearized effective Hamiltonian becomes:
\begin{equation}
\label{SH_Hamiltonian-effective}
    \hat{H}_{SH}^{eff} = \sum_k\left(E_{k\ua}\hat{n}_{k\ua} + E_{k\da}\hat{n}_{k\da} \right) - VU n_\ua n_\da.
\end{equation}
Here, $n_\sigma = N_{\sigma}/V$ is the density of
electrons with spin projection along $z$-axis given by
$\sigma\in\{\ua,\da\}$ ($N_{\sigma}=\sum_k \langle \hat{n}_{k\sigma}
\rangle$ being the total number of electrons with spin
$\sigma$).
The one-particle electron energies in this effective model are
obtained by shifting the free-electron energies $\varepsilon_k$ by
an amount depending on the particle's spin:
\begin{eqnarray}
\label{SH-energies}
    E_{k\ua} = \varepsilon_k + Un_\da, \nonumber \\ 
    E_{k\da} = \varepsilon_k + Un_\ua.              
\end{eqnarray}
Since the one-particle energy modes are decoupled, the overall ground state is
obtained by filling the electrons up to the Fermi level
$\varepsilon_F$:
\begin{equation}
\label{SH-ground_state}
    |g\rangle = \otimes_{k\leq k_{F\ua}} \hat{c}^\dag_{k\ua}\otimes_{k\leq k_{F\da}} \hat{c}^\dag_{k\da}|0\rangle,
\end{equation}
where $|0\rangle$ represents the vacuum state with no electrons and $k_{F\ua}$ the maximal
value of the momentum for spin up electrons, given by
$E_{k_{F\ua}} = \varepsilon_F$ (and analogously
for $k_{F\da}$) \footnote{For the proof of the relation $\varepsilon_F = E_{k_{F\ua}} =
E_{k_{F\da}}$ see Appendix $1$.}. Note that, due
to the different dispersion formulas (\ref{SH-energies}) for particles
with spin up and spin down, in general the values of $k_{F\ua}$ and
$k_{F\da}$ that minimize the ground state energy, and therefore
define the state (\ref{SH-ground_state}) itself, are different and
consequently the number of up and down electrons will be different. This
is precisely the reason for the existence of magnetism in this
model. As soon as the energy of the ``biased'' (magnetic) state, for
which for example $k_{F\ua}>k_{F\da}$, becomes lower than the energy
of the ``balanced'' (paramagnetic) state ($k_{F\ua}=k_{F\da}$), a
magnetic phase transition will occur. Obviously, for reasons of symmetry,
the magnetic state can be reversed with the
$k_{F\ua}>k_{F\da}$ and $k_{F\ua}<k_{F\da}$ cases having the same
energy, in the absence of an external symmetry breaking field $\vec{H}$. The qualitative picture of the emergence of magnetic
features can be seen already from looking at the original
Hamiltonian (\ref{SH_Hamiltonian-original}): in the $U\rightarrow 0$
limit, when the Coulomb interaction is negligible, the Hamiltonian
represents a system of free electrons that exhibits no magnetic
order (all Bloch states are doubly occupied); in the opposite
$U\rightarrow\infty$ limit, the second term of the Hamiltonian
becomes the dominant one and is minimized by one of two possible
states for which either $N_{\ua}=0$ or $N_{\da}=0$.

The quantitative analysis of the zero-temperature critical behavior
of the effective Hamiltonian (\ref{SH_Hamiltonian-effective}) can be
done by looking at the divergence of the magnetic susceptibility, using the one-particle
energy dispersion relations (\ref{SH-energies}). At $T=0$ it leads to the well known Stoner
criterion for the emergence of magnetism \cite{stoner}:
\begin{equation}
\label{SH_criterion}
    D_FU_c=1,
\end{equation}
where $U_c$ is the critical value of the coupling constant above
which the system is in a magnetic phase and $D_F=D(\varepsilon_F)$
is the density of states around the Fermi energy. From this, one can
obtain (see Appendix $1$) the critical value $U_c$ of the coupling
constant \footnote{In the following, we discuss the case of a
$3$-dimensional spatial lattice. This treatment could be
easily generalized to a different lattice dimensionality.}:
\begin{equation}
\label{U_c}
    U_c = \frac{4}{3} \ \varepsilon_F \frac{V}{N},
\end{equation}
where $N$ is the total number of electrons, $V$ is the volume of the system and $m$ is the electron mass.

Alternatively, the above result can be derived by minimizing the
overall ground state energy, thus obtaining the explicit
dependencies $k_{F\ua}=k_{F\ua}(U)$ and $k_{F\da}=k_{F\da}(U)$ which
determine the ground state (\ref{SH-ground_state}) and $U_c$ in
particular (the maximum value of $U$ for which
$k_{F\ua}=k_{F\da}$) \footnote{Strictly speaking, for $U>U_c$ we
speak of multi (two-fold) valued functions $k_{F\ua}(U)$ and
$k_{F\da}(U)$.}. The total number of electrons in the system is given by $N = N_\ua
+ N_\da$, and thus:
\begin{equation}
\label{SH-eq1}
    \frac{N}{V}=\frac{1}{6\pi^2}[(k_{F\ua})^3 + (k_{F\da})^3],
\end{equation}
since the up and down electrons occupy spheres of radius $k_{F\ua}$
and $k_{F\da}$, with volumes $V(k_{F\ua})$ and $V(k_{F\da})$, in
momentum space, with a density of states $V/(2\pi)^3$, as follows
from the periodic boundary conditions for the Bloch functions.
From $E_{k_{F\ua}} =
E_{k_{F\da}}$ (see Appendix $1$), using the energy
dispersion formulas (\ref{SH-energies}), we obtain the second
equation that determines the Fermi momenta $k_{F\ua}$ and $k_{F\da}$ (with $\alpha = \frac{3}{2}k_FU_c$):
\begin{equation}
\label{SH-eq2}
 (k_{F\ua} - k_{F\da})[(k_{F\ua} + k_{F\da}) - \frac{U}{\alpha}(k_{F\ua}^2 + k_{F\ua}k_{F\da} + k_{F\da}^2 )] = 0.
\end{equation}

We see that the above two equations always have the trivial solution
$k_{F\ua}=k_{F\da}\equiv k_F$, which from (\ref{SH-eq1}) is $k_F =
\left(3\pi^2\frac{N}{V}\right)^{1/3}$. Yet, it is not necessarily
the only possible solution, as  the quadratic term in equation
(\ref{SH-eq2}) can also be satisfied, $(k_{F\ua} + k_{F\da}) - \frac{U}{\alpha}(k_{F\ua}^2 + k_{F\ua}k_{F\da} + k_{F\da}^2 ) = 0$. It turns out that precisely for $U > U_c$ this term
has non-trivial, ``non-balanced'' solutions that are energetically
more favorable than the ``balanced'' one and that give rise to
magnetic order (see Appendix $1$).

Thus, equations (\ref{SH-eq1}) and (\ref{SH-eq2}) define
$k_{F\ua}=k_{F\ua}(U)$ and $k_{F\da}=k_{F\da}(U)$ functions which,
via (\ref{SH-ground_state}), give the ground state as a function of
the external parameter $U$, $|g\rangle = |g(U)\rangle$. This enables
us to analyze the fidelity between two ground states
$|g\rangle\equiv|g(U)\rangle$ and $|\tilde{g}\rangle=|g(U+\delta
U)\rangle$ in two close parameter points $U$ and $U+\delta U$. The
fidelity is then $F(|g\rangle\langle
g|,|\tilde{g}\rangle\langle\tilde{g}|) = |\langle
g|\tilde{g}\rangle|$, and for $U<U_c$ (and $\delta U$ sufficiently
small, i.e. $\delta U <U_c-U$) we see that the fidelity is identical
to $1$. This is a simple consequence of our mean-field approximation
based on a simplified description in terms of single particle energy
states. On the magnetic side of the phase diagram, the ground states
are indeed different from each other, which follows from the
relation $k_{F\ua}(U)\neq k_{F\da}(U)$. In fact, from equation
(\ref{SH-ground_state}) it follows that \emph{any} two ground states
with different numbers $N_{\ua}$ and $N_{\da}$ are \emph{orthogonal}
to each other. Since this is precisely the case in the thermodynamic
limit, the fidelity between any two different ground states (in two
different parameter points) is identically equal to zero. This is
the famous Anderson orthogonality catastrophe, discussed
in more detail in \cite{us}. For systems with infinitely many
degrees of freedom, such as those taken in the thermodynamic limit
are, every two ground states are generally orthogonal to each other.
Therefore, in order to infer the points of criticality, we are forced to either analyze the finite-size scaling behavior (see \cite{us}), or to introduce the fidelity per lattice site and work directly in the thermodynamic limit (see \cite{zhou-1}, \cite{zhou-2} and \cite{vidal}).


In our case though, even for finite systems the ground state is
discontinuous at the point of criticality. This is due to the fact
that the unperturbed and the symmetry-breaking perturbation
Hamiltonian commute with each other, which results in a first-order
quantum phase transition at the point of level-crossing between the
ground and the first excited state. In other words, in the case of
finite systems ($N,V < \infty$), the small enough changes of the
parameter $U\rightarrow U+\delta U$ will result in small changes of
volumes $V(k_{F\ua})\rightarrow V(k_{F\ua})+\delta V(k_{F\ua})$ such
that $|\delta V(k_{F\ua})|<V_0=(2\pi)^3/V$ ($V_0$ is the volume
occupied by each one-particle state in momentum space):
infinitesimal changes of the volumes $V(k_{F\ua})$ and $V(k_{F\da})$ are
small enough to cause a change in the numbers $N_{\ua}$ and $N_{\da}$,
and thus in the ground state. Therefore, for finite systems, the
fidelity between two ground states is either one or zero -- it is
not a continuous function of and its rate of change can not be analyzed
directly.

\begin{figure}[ht]
\centering
\includegraphics[width=5.5cm,height=7.0cm,angle=-90]{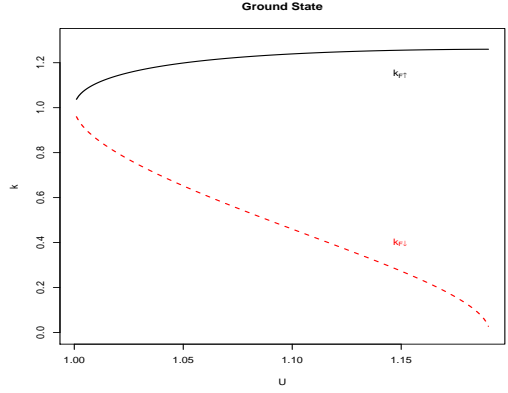}
\caption{(color online) The explicit dependencies $k_{F\ua}=k_{F\ua}(U)$ (black, solid line) and $k_{F\da}=k_{F\da}(U)$ (red, dashed line)
determining the ground state (\ref{SH-ground_state}). Note that at the point of QPT, given by $U_c=1$, the two derivatives $\frac{dk_{F\ua}}{d U}$ and $\frac{dk_{F\da}}{d U}$ become infinite, marking the non-analyticity in the ground state fidelity. The plot is given in rescaled quantities $U\rightarrow D_FU$ and $k_{F\sigma}\rightarrow k_{F\sigma}/k_F$.} \label{Stoner-GS Figure}
\end{figure}

Yet, we can use the rate of change of $V(k_{F\ua})$, the derivative $\frac{d V(k_{F\ua})}{d U}$, to quantify the change of fidelity itself
(note that, due to the fixed total number of electrons $N$, we have
that $\frac{d V(k_{F\da})}{d U} = - \frac{d V(k_{F\ua})}{d U}$).
Lengthy, but elementary algebra (see Appendix $1$) shows that precisely at $U=U_c$, the derivative
$\frac{d V(k_{F\ua})}{d U}$ diverges to infinity, thus marking the \emph{macroscopic}
distinguishability between the states from paramagnetic and magnetic
phase (see Fig. \ref{Stoner-GS Figure}).


Next, we discuss the general case of $T\neq 0$ phase transitions.
First, we transform the Hamiltonian $\hat{H}_{SH}^{eff}$ to a form
with the explicit symmetry breaking term that drives the phase
transition. Using $M=(N_\ua - N_\da)/2=\langle\hat{S}^z\rangle=\sum_k\langle\hat{S}_k^z\rangle$, with $N_\sigma =
\langle\hat{N}_\sigma\rangle=\sum_k\langle\hat{n}_{k\sigma}\rangle$, $\hat{n}_k=\hat{n}_{k\ua}+\hat{n}_{k\da}$
and
$\hat{S}_k^z=\frac{1}{2}(\hat{n}_{k\ua}-\hat{n}_{k\da})=\hat{\Psi}^\dag\frac{\sigma^z}{2}\hat{\Psi}$,
where $\hat{\Psi}^\dag_k = [\hat{c}_{k\ua}^\dag  \ \hat{c}_{k\da}^\dag ]$
and $\sigma^z$ is the $z$-component of the vector
$\vec{\sigma}$ of Pauli matrices, we get:
\begin{equation}
\hat{H}_{SH}^{eff}=\sum_k\left[\left(\varepsilon_k+\frac{UN}{2V}\right)\hat{n}_k-2\frac{UM}{V}\hat{S}_k^z\right]-\frac{U}{V}\left(\frac{N^2}{4}-M^2\right).
\end{equation}

In thermal equilibrium, the state of the system is given by
$\hat{\rho} = \frac{1}{Z}e^{-\beta(\hat{H}_{SH}^{eff}-\mu\hat{N})}$,
where $Z=\mbox{Tr}[e^{-\beta(\hat{H}_{SH}^{eff}-\mu\hat{N})}]$ is
the grand canonical partition function, $\beta = 1/(k_BT)$, with $k_B$
being the Boltzmann constant and $T$ the absolute temperature and
$\mu=\mu(T)$ is the chemical potential. Using the above expression
for the Hamiltonian $\hat{H}_{SH}^{eff}$, we can write:
\begin{equation}
-\beta(\hat{H}_{SH}^{eff}-\mu\hat{N})=\sum_k\left(\alpha_k\hat{n}_k+h_z\hat{S}_k^z\right)+C.
\end{equation}
Here, $\alpha_k=-\beta E_k$ (with $E_k=\bar{\varepsilon}_k+\frac{UN}{2V}$ and
$\bar{\varepsilon}_k=\varepsilon_k-\mu$), $h_z=2\beta
\frac{U}{V}M$ and $C=\beta\frac{U}{V}(\frac{N^2}{4}-M)$.
Note that the coefficients $\alpha_k$, $h_z$ and $C$ are functions
of both the coupling constant $U$ and, through $\beta$ and the
chemical potential $\mu=\mu(T)$, of the temperature $T$ as well, so that
the ``generalized'' parameter is $q=(T,U)$. Using the obvious
commutation relations
$[\hat{n}_k,\hat{n}_{k^\prime}]=[\hat{n}_k,\hat{S}_{k^\prime}^z]=[\hat{S}_k^z,\hat{S}_{k\prime}^z]=0$
(for $k\neq k^\prime$), the equilibrium state can be expressed as:
\begin{eqnarray}
\label{SH-state}
\hat{\rho} \! = \! \frac{1}{Z}e^{-\beta(\hat{H}_{SH}^{eff}-\mu\hat{N})} \! = \! \frac{e^{\sum_k\left(\alpha_k\hat{n}_k+h_z\hat{S}_k^z\right)+C}}{\mbox{Tr}\left[e^{\sum_k\left(\alpha_k\hat{n}_k+h_z\hat{S}_k^z\right)+C} \right]} =  \frac{\prod_k\left(e^{\alpha_k\hat{n}_k}e^{h_z\hat{S}_k^z}\right)}{\prod_k\mbox{Tr}\left[e^{\alpha_k\hat{n}_k}e^{h_z\hat{S}_k^z}\right]}.
\end{eqnarray}

We next choose two parameter points $q_a=(T_a,U_a)$ and
$q_b=(T_b,U_b)$ defining the Hamiltonians
$\hat{H}_a=\hat{H}_{SH}^{eff}(U_a)$ and
$\hat{H}_b=\hat{H}_{SH}^{eff}(U_b)$ and the corresponding
equilibrium states $\hat{\rho}_a=\hat{\rho}(q_a)$ and
$\hat{\rho}_b=\hat{\rho}(q_b)$ respectively. The fidelity between
the two states is then given by:
\begin{eqnarray}
\label{fidelity-SH}
    F(\hat{\rho}_a, \hat{\rho}_b) = \mbox{Tr} [(\hat{\rho}_a^{1/2}\hat{\rho}_b\hat{\rho}_a^{1/2})^{1/2}] = \mbox{Tr} [\sqrt{\hat{\rho}_a\hat{\rho}_b}]  = \frac{\mbox{Tr}\left[e^{\frac{\beta_a\hat{H}_a+\beta_b\hat{H}_b}{2}}\right]}{\sqrt{Z(\hat{H}_a)Z(\hat{H}_b)}}.
\end{eqnarray}
Using (\ref{SH-state}) and the expression
$\mbox{Tr}[e^{(\alpha_k\hat{n}_k+h_z\hat{S}^z_k)}]=2e^{\alpha_k}(\cosh
\alpha_k + \cosh \frac{h_z}{2})$ (for the proof, see Appendix $2$),
the fidelity between two different equilibrium states $\hat{\rho}_a$
and $\hat{\rho}_b$ finally becomes:
\begin{equation}
\label{SH_fidelity-final}
    F(\hat{\rho}_a, \hat{\rho}_b)=\prod_k\frac{\cosh \bar{\alpha}_k + \cosh \frac{\bar{h}_z}{2}}{\sqrt{\left[ \cosh (\bar{\alpha}_k+\frac{\Delta\alpha_k}{2}) + \cosh (\frac{\bar{h}_z}{2}+\frac{\Delta h_z}{2}) \right] \left[ \cosh (\bar{\alpha}_k-\frac{\Delta\alpha_k}{2}) + \cosh (\frac{\bar{h}_z}{2}-\frac{\Delta h_z}{2}) \right]}},
\end{equation}
with $\bar{\alpha}_k=(\alpha_k(q_a)+\alpha_k(q_b))/2$,
$\Delta\alpha_k=\alpha_k(q_a)-\alpha_k(q_b)$, and similarly for
$\bar{h}_z$ and $\Delta h_z$. If we choose the two points to be
close to each other, $\Delta\alpha_k << 1$ and $\Delta h_z << 1$, then
the fidelity can be seen as a function of $\bar{\alpha}_k$ and $\bar{h}_z$, with a fixed parameter difference.

\begin{figure}[h]
\centering
\includegraphics[width=6.5cm,height=7.0cm,angle=-90]{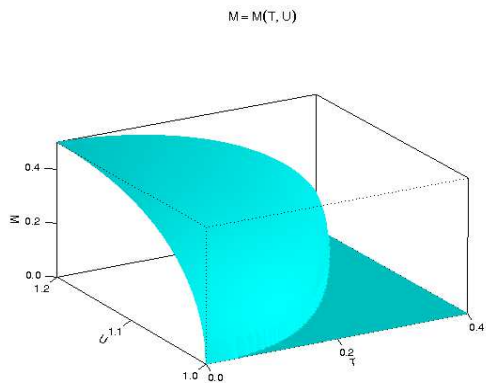}
\caption{(color online) Magnetization $M=M(T,U)$ as a function of the temperature $T$ and the coupling constant $U$. The plot is given in rescaled quantities $T\rightarrow k_BT$, $U\rightarrow D_FU$ and $M\rightarrow M/N$.} \label{Stoner-Magnetization Figure}
\end{figure}

\begin{figure}[h]
\centering
\includegraphics[width=5.0cm,height=7.0cm,angle=-90]{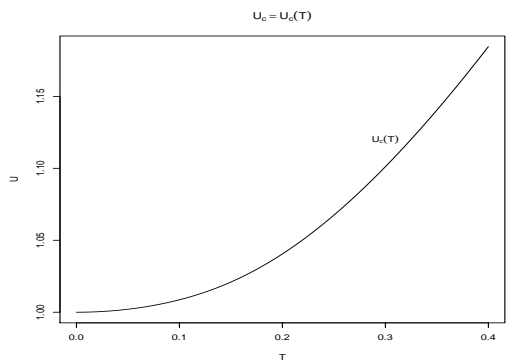}
\caption{The critical line of the magnetic phase transition $U_c=U_c(T)$. The plot is given in rescaled quantities $T\rightarrow k_BT$ and $U\rightarrow D_FU$.} \label{Stoner-Uc Figure}
\end{figure}

\begin{figure}[h]
\centering
\includegraphics[width=6.5cm,height=7.0cm,angle=-90]{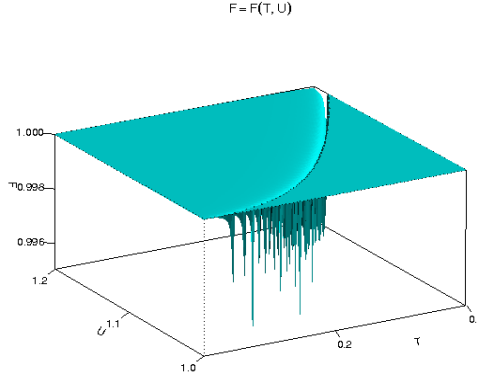}
\caption{(color online) The fidelity $F=F(T,U)$. The plot is given in rescaled quantities $T\rightarrow k_BT$ and $U\rightarrow D_FU$, with $\delta T =0, \delta U = 2\times10^{-3}$.} \label{Stoner-Fidelity Figure}
\end{figure}

In order to evaluate the fidelity (\ref{SH_fidelity-final}), we need
to determine the magnetization $M=M(T,U)$ and the chemical potential
$\mu=\mu(T,U)$, given by the pair of self consistent integral
equations:
\begin{eqnarray}
    N & = & VD_F\int_0^{+\infty}d\varepsilon \sqrt{\frac{\varepsilon}{\varepsilon_F}}\left[ f(E_{k\ua}) + f(E_{k\da})\right],
    \nonumber \\
    M & = & VD_F\int_0^{+\infty}d\varepsilon \sqrt{\frac{\varepsilon}{\varepsilon_F}}\frac{1}{2}\left[ f(E_{k\ua}) - f(E_{k\da})\right],
\end{eqnarray}
where $f(E_{k\sigma})=[\exp (\beta E_{k\sigma})+1]^{-1}$ is the
usual Fermi distribution. We used the subroutine hybrd.f from
MINPACK \cite{programme} to solve the above system numerically. In
the $T\rightarrow 0$ limit, the above system reduces to equations
(\ref{SH-eq1}) and (\ref{SH-eq2}) that determine the $T=0$ ground
state. The result for the magnetization is given on Fig.
\ref{Stoner-Magnetization Figure}. The line of the phase transition
$U_c=U_c(T)$ is clearly marked, and is plotted on Fig.
\ref{Stoner-Uc Figure}. Finally, using the numerical results for $M$
and $\mu$, we obtain the fidelity, depicted on Fig.
\ref{Stoner-Fidelity Figure}. We clearly see the same line of the
phase transition as the line of a sudden drop of $F$. Note that all
the plots are given in rescaled quantities $T\rightarrow k_BT$,
$U\rightarrow D_FU$ and $M\rightarrow M/N$, with $\delta T =0,
\delta U = 2\times10^{-3}$. We have evaluated the fidelity for
$\delta T =2\times10^{-3}$ and $\delta U = 0$, as well as for $\delta
T = \delta U = 2\times10^{-3}$ and the results are qualitatively the
same.


\section{The general case of mutually commuting Hamiltonians}

The above singular behavior of the fidelity indeed marks the
regions of phase transitions, not only for the case of
the Stoner-Hubbard model but also for the broad class of systems given
by a set of mutually commuting Hamiltonians $\hat{H}(q)$ and whose
critical behavior is explained by the LGW theory, as can be seen by the following analysis. Using the mean Hamiltonian
$\hat{\bar{H}}=(\hat{H}_a+\hat{H}_b)/2$ and the difference
$\Delta\hat{H}=\hat{H}_a-\hat{H}_b$, from equation
(\ref{fidelity-SH}) the fidelity can be written as:
\begin{equation}\label{fid-gen_comm}
    F(\hat{\rho}_a, \hat{\rho}_b) = \frac{Z(\hat{\bar{H}})}{\sqrt{Z(\hat{\bar{H}}+\frac{\Delta\hat{H}}{2})Z(\hat{\bar{H}}-\frac{\Delta\hat{H}}{2})}}.
\end{equation}
Within the LGW theory, PTs occur as a consequence of the emergence
of a symmetry-breaking term in the Hamiltonian, given by an operator
$\hat{S}$. Thus, we can write the overall Hamiltonian as a sum of
unperturbed and symmetry-breaking terms, $\hat{H}=\hat{H}_0 -
h\hat{S}$, with $h=h(q)$. Using \footnote{For reasons of simplicity,
we use formulas for canonical rather than grand canonical ensemble.
Introducing chemical potential results only in a ``shift'' of the
one-particle energies from the unperturbed Hamiltonian $\hat{H}_0$.}
$Z(\hat{H})=\mbox{Tr}[e^{-\beta(\hat{H}_0-h\hat{S})}]$ (note the
implicit dependence of partition function $Z$ on temperature $T$,
through $\beta$), for the first and the second derivative we obtain
$\frac{\partial \ln Z}{\partial h} = \beta
\mbox{Tr}[\hat{\rho}(\hat{H})\hat{S}] = \beta \langle\hat{S}\rangle
= \beta M$ and $\frac{\partial^2}{\partial h^2}(\ln Z) = \beta
\chi$. Here, $M=M(h)$ and
$\chi=\chi(h)=\beta(\langle\hat{S}^2\rangle -
\langle\hat{S}\rangle^2)$ are the generalized ``magnetization'' and
``susceptibility'' respectively. Thus, for the logarithm of the
partition function (free energy) we obtain $\ln
Z(\hat{H})|_{h+\Delta h} = \ln Z(\hat{H})|_{h}+\beta M(h)\Delta h +
\frac{1}{2}\beta \chi(h) \Delta h^2 + o(\Delta h^2)$. Finally, in
the vicinity of the points of parameter space where the
self-consistent field $h$ vanishes, the fidelity
(\ref{fid-gen_comm}) reads as:
\begin{equation}
\label{commuting_fidelity}
    F|_{h=\Delta h} \simeq e^{- \frac{1}{2}\beta\chi(0)\Delta h^2}.
\end{equation}
According to the LGW theory, at the phase transition, the
zeroth-field ``susceptibility'' $\chi(0)$ becomes non-analytic and
diverges. Thus it follows that the fidelity $F$ will itself become
nonanalytic and experience a sudden drop.

When interested in the system's behavior at PTs, we usually simplify
the problem by taking the unperturbed Hamiltonian to be constant.
Yet, in general, the unperturbed Hamiltonian is also a function of
the parameters, $\hat{H}_0=\hat{H}_0(q)$. This gives the correction
to the above formula for the fidelity which, introducing
$\hat{\bar{H}} = \hat{\bar{H}}_0 - \bar{h}\hat{S}$ and
$\Delta\hat{H} = \Delta\hat{H}_0 - \Delta h \hat{S}$, reads as:
\begin{equation}
    F|_{h=\Delta h} \simeq \frac{Z(\hat{\bar{H}}_0)}{\left[Z(\hat{\bar{H}}_0 + \frac{\Delta\hat{H}_0}{2}) Z(\hat{\bar{H}}_0 - \frac{\Delta\hat{H}_0}{2})\right]^{1/2}} \ \ e^{- \frac{1}{2}\beta\chi(0)\Delta h^2}.
\end{equation}
The ``correction'' term in the above product is
responsible for ``short-range'' local correlations, while the second
one quantifies the global ``long-range'' correlations giving rise to
macroscopic phase distinguishability. Note though that even within a
single phase, a system can be in different macroscopically distinguishable
states -- phase distinguishability is not the only form of
macroscopic distinguishability. Yet, it is in some sense the
``extreme'' version of it, which clearly affects the behavior of
fidelity.


Finally, we note that in the above discussion we focused on PTs
driven by the local order parameter $\hat{S}$. Therefore, we
analyzed the Taylor expansion of $F$ with respect to $\Delta h$
deviations only, which at the second order are given by the
generalized susceptibility $\chi(h)$. In the general case,
considering the temperature deviations as well, one would include
additional terms involving the specific heat $C=C(q)$, again
resulting in a singular behavior of the fidelity.


\section{BCS Superconductivity}


Next, we discuss the BCS theory for superconductivity \cite{stoner}, providing us
with an example of a model with mutually non-commuting Hamiltonians.
The one-electron Bloch momentum modes are given by the fermionic
anti-commuting operators $\hat{c}_{k\sigma}$ (label
$\sigma\in\{\ua,\da\}$ represents spins with projections up and down
along, say $z$-axis), with the one-particle kinetic energies taken
to be, again for simplicity, $\varepsilon_k=\hbar^2k^2/(2m)$. The
BCS superconducting Hamiltonian that represents the sum of
one-particle kinetic and Cooper-pair interaction energies can be
written in the following way ($V_{k^\prime k}=V^\ast_{kk^\prime}$
are the coupling constants):
\begin{equation}
    \hat{H}_{BCS} = \sum_{k\sigma} \varepsilon_k \hat{c}^\dag_{k\sigma}\hat{c}_{k\sigma} + \sum_{kk^\prime} V_{kk^\prime}\hat{c}^\dag_{k^\prime\ua}\hat{c}^\dag_{-k^\prime\da}\hat{c}_{-k\da}\hat{c}_{k,\ua}.
\end{equation}
By $\hat{n}_{k\sigma} = \hat{c}^\dag_{k\sigma}\hat{c}_{k\sigma}$ we
denote the one-particle number operators, while by
$\hat{b}^\dag_k=\hat{c}^\dag_{k\ua}\hat{c}^\dag_{-k\da}$ and
$\hat{b}_k=\hat{c}_{-k\da}\hat{c}_{k\ua}$ we define the Cooper-pair
creation and annihilation operators respectively. Analogously to the
previous case, using $\hat{b}_k =
\langle\hat{b}_k\rangle+\delta\hat{b}_k$ and neglecting the term
quadratic in the fluctuations, we obtain the effective mean-field
BCS Hamiltonian:
\begin{equation}
\label{bcs-hamiltonian}
    \hat{H}_{BCS}^{eff} = \sum_{k} \varepsilon_k(\hat{n}_{k\ua}+\hat{n}_{-k\da}) - \sum_k(\Delta_k\hat{b}^\dag_k + \Delta^\ast_k\hat{b}_k - \Delta^\ast_k b_k),
\end{equation}
with $\Delta_k=-\sum_{k^\prime} V_{kk^\prime}b_{k^\prime}$ and
$b_{k}=\langle\hat{b}_k\rangle$. We will use the usual assumption
that the lattice-mediated pairing interaction is constant and
non-vanishing between electrons around the Fermi level only, i.e.
$V_{kk^\prime}=-V$ for $|\bar{\varepsilon}_k|$ and
$|\bar{\varepsilon}_{k^\prime}|<\hbar\omega_D$, and zero otherwise
($\omega_D$ is the Debye frequency). Using the Nambu operators \cite{stoner} $\hat{\vec{T}}_k =
\hat{\psi}^\dag_k \frac{\vec{\sigma}}{2} \hat{\psi}_k$, where
$\hat{\psi}^\dag_k = [\hat{c}_{k\ua}^\dag  \ \hat{c}_{-k\da} ]$ and
$\vec{\sigma}$ is the vector of Pauli matrices \footnote{Note the
similarity, and the difference, to the previously introduced operators
$\hat{\Psi}_k$.}, the operators are given by $\hat{T}_k^{+} = \hat{b}^\dag_k$,
$\hat{T}_k^{-} = \hat{b}_k$ and $2\hat{T}_k^0 + 1 = (\hat{n}_{k\ua}+\hat{n}_{-k\da})$
and form a $\mbox{su}(2)$ algebra (see the Appendix $3$).
Using this notation, the Hamiltonian takes the form:
\begin{equation}
    \hat{H}_{BCS}^{eff} = \sum_k (2\varepsilon_k\hat{T}_k^0 - \Delta_k\hat{T}_k^{+} - \Delta^\ast_k\hat{T}_k^{-}) + \sum_k(\varepsilon_k + \Delta^\ast_k b_k).
\end{equation}

As before, the thermal equilibrium state is given by $\hat{\rho}
= \frac{1}{Z}e^{-\beta(\hat{H}_{BCS}^{eff}-\mu\hat{N})}$ and, using
the above expression for the Hamiltonian $\hat{H}_{BCS}^{eff}$, we
can write:
\begin{equation}
    -\beta(\hat{H}_{BCS}^{eff}-\mu\hat{N}) = \sum_k\vec{\tilde{h}}_k \hat{\vec{T}}_k + K,
\end{equation}
where $\vec{\tilde{h}}_k =
(\tilde{h}_k^{+},\tilde{h}_k^{-},\tilde{h}_k^0) =
(2\beta\Delta^\ast_k,2\beta\Delta_k,-2\beta\bar{\varepsilon}_k)$,
$\hat{\vec{T}}_k = (\hat{T}_k^{+},\hat{T}_k^{-},\hat{T}_k^0)$,
$K=-\beta\sum_k (\bar{\varepsilon}_k + \Delta^\ast_k b_k)$ and
$\bar{\varepsilon}_k=\varepsilon_k-\mu(T)$. The norms of the vectors
$\vec{\tilde{h}}_k$ are given by $\tilde{h}_k = 2\beta
E_k$, with $E_k =
\sqrt{\bar{\varepsilon}_k^2+\left|\Delta_k\right|^2}$. Similarly
to what we had before, the coefficients $\vec{\tilde{h}}_k =
\vec{\tilde{h}}_k(T,V)$ are functions of both the coupling
constant $V$ and the temperature $T$,
through the gap parameters $\Delta_k = \Delta_k(T,V)$ and the
chemical potential $\mu=\mu(T)$. Thus,
we can talk of the ``generalized'' parameter $q=(T,V)$. Since
$[\hat{\vec{T}}_k,\hat{\vec{T}}_{k^\prime}]=0$, for $k \neq
k^\prime$, we have:
\begin{equation}
    \hat{\rho} = \frac{1}{Z}e^{-\beta(\hat{H}_{BCS}^{eff}-\mu\hat{N})} = \frac{e^{\sum_k\vec{\tilde{h}}_k \hat{\vec{T}}_k + K}}{\mbox{Tr}[e^{\sum_k\vec{\tilde{h}}_k \hat{\vec{T}}_k + K}]} =  \frac{\prod_k e^{\vec{´\tilde{h}}_k \hat{\vec{T}}_k}}{\prod_k\mbox{Tr}[e^{\vec{\tilde{h}}_k \hat{\vec{T}}_k}]}.
\end{equation}

We wish to evaluate the fidelity between two thermal states
$\hat{\rho}_a$ and $\hat{\rho}_b$, given for two different parameter
points $q_a=(T_a,V_a)$ and $q_b=(T_b,V_b)$. Using definition
(\ref{fidelity-def}) and $\vec{a}_k = \vec{\tilde{h}}_k(q_a)$
and $\vec{b}_k = \vec{\tilde{h}}_k(q_b)$, we have:
\begin{eqnarray}
    F(\hat{\rho}_a, \hat{\rho}_b) = \mbox{Tr} [(\hat{\rho}_a^{1/2}\hat{\rho}_b\hat{\rho}_a^{1/2})^{1/2}] = \frac{\mbox{Tr}[(\prod_k e^{\frac{\vec{a}_k}{2} \hat{\vec{T}}_k} e^{\vec{b}_k \hat{\vec{T}}_k} e^{\frac{\vec{a}_k}{2} \hat{\vec{T}}_k})^{1/2}]}{\prod_k (\mbox{Tr}[e^{\vec{a}_k \hat{\vec{T}}_k}] \mbox{Tr}[e^{\vec{b}_k \hat{\vec{T}}_k}])^{1/2}}.
\end{eqnarray}

As for every $k$ the operators $\hat{\vec{T}}_k$ form a
$\mbox{su}(2)$ algebra, and therefore by exponentiation define a Lie
group, we can write $e^{\frac{\vec{a}_k}{2} \hat{\vec{T}}_k}
e^{\vec{b}_k \hat{\vec{T}}_k} e^{\frac{\vec{a}_k}{2}
\hat{\vec{T}}_k} = e^{2\vec{c}_k \hat{\vec{T}}_k}$. Also (see
Appendix $3$), we have that $\mbox{Tr}[e^{\vec{a}_k
\hat{\vec{T}}_k}] = 2(1+\cosh\frac{a_k}{2})$. Therefore, we finally
have (see Appendix $3$) that $F(\hat{\rho}_a, \hat{\rho}_b) = \prod_k F_k(\hat{\rho}_a,
\hat{\rho}_b)$ with \footnote{We use the well known identity
$\cosh\frac{c_k}{2} = \sqrt{\frac{1}{2}(1+\cosh c_k)}$.}:
\begin{eqnarray}
\label{BCS-fidelity}
   F_k(\hat{\rho}_a, \hat{\rho}_b) \!\!\! & = & \!\!\! \frac{\mbox{Tr}[e^{\vec{c}_k \hat{\vec{T}}_k}]}{\sqrt{\mbox{Tr}[e^{\vec{a}_k \hat{\vec{T}}_k}]\mbox{Tr}[e^{\vec{b}_k \hat{\vec{T}}_k}]}} = \frac{1+\sqrt{\frac{1}{2}(1+\cosh c_k)}}{\sqrt{(1+\cosh\frac{a_k}{2})(1+\cosh\frac{b_k}{2})}}, \ \ \ \ \ \ \mbox{and} \\
\label{cosh-explicit}
  \cosh c_k \!\!\! & = & \!\!\! \cosh(\beta^aE_k^a)\cosh(\beta^bE_k^b)\left\{1 + \tanh(\beta^aE_k^a)\tanh(\beta^bE_k^b) \frac{\bar{\varepsilon}_k^a\bar{\varepsilon}_k^b + \mbox{Re}[\Delta_k^a(\Delta_k^b)^\ast]}{E_k^aE_k^b}\right\}.
\end{eqnarray}
Here, we used the relation $\bar{\varepsilon}_k^a =
\varepsilon_k^a-\mu^a$ and $a_k=2\beta^aE_k^a$, $E_k^a =
\sqrt{(\bar{\varepsilon}_k^a)^2+\left|\Delta_k^a\right|^2}$, so that
$\cosh\frac{a_k}{2} = \cosh(\beta^aE_k^a)$ (and analogously for
$q_b=(V_b,T_b)$). Note the explicit dependence of all the quantities
on the temperature and the coupling strength, given through the
superscripts $a$ and $b$, denoting two parameter points
$q_a=(T_a,V_a)$ and $q_b=(T_b,V_b)$.

Assuming that the chemical
potential is also constant ($\mu=\varepsilon_F$) in the region of
interest, where the phase transition takes place, the gap parameter
reduces to $\Delta_k=\Delta$, for
$|\bar{\varepsilon}|<\hbar\omega_D$ (and zero otherwise). Thus, the
self consistent equation for the gap
$\Delta_k=-\sum_{k^\prime}V_{kk^\prime}\frac{1-2f(E_{k^\prime})}{2E_{k^\prime}}\Delta_{k^\prime}$
reads as ($1-2f(E_{k^\prime})=\tanh\frac{\beta E_{k^\prime}}{2}$):
\begin{equation}
    1=D_FV\int_{-\hbar\omega_D}^{\hbar\omega_D}d\varepsilon \  \frac{\tanh\frac{\beta}{2}\sqrt{\varepsilon^2+\Delta^2(T,V)}}{2\sqrt{\varepsilon^2+\Delta^2(T,V)}},
\end{equation}
with $f(E)$ being the Fermi distribution.

In the $T\rightarrow 0$ limit, we obtain the expression for the
ground state fidelity $F(|g_a\rangle\langle g_a|,|g_b\rangle\langle
g_b|)=|\langle g_a|g_b\rangle|$. At zero temperature, the chemical
potential is equal to the Fermi energy $\varepsilon_F$,
$\mu(T=0)=\varepsilon_F$. Further, the gap equation reduces to
$\Delta(V) = \hbar\omega_D/(\sinh\frac{2}{D_FV}) \simeq
2\hbar\omega_D\exp(-2/D_FV)$, where $D_F$ is the density of states
around the Fermi level. The zero temperature ground state fidelity
is $F(|g_a\rangle\langle g_a|,|g_b\rangle\langle
g_a|)=\prod_k\frac{1}{\sqrt{2}}\left(1+\frac{(\varepsilon_k-\varepsilon_F)^2+\Delta(V_a)\Delta(V_b)}{\sqrt{(\varepsilon_k-\varepsilon_F)^2+\Delta(V_a)^2}\sqrt{(\varepsilon_k-\varepsilon_F)^2+\Delta(V_b)^2}}\right)^{1/2}$,
which matches the expression one obtains for $T=0$. In other words,
we see that, as in the case of the Stoner-Hubbard model, the point
of criticality of the $T=0$ QPT can be inferred from the mixed state
fidelity between the thermal states. The phase transition from
superconductor to normal metal happens at $V=0$. Thus, for
$V_a\rightarrow 0_+$ and $V_b=V_a+\delta V \rightarrow \delta V>0$,
the fidelity between the corresponding ground states of Fermi sea
$|g_F\rangle$ and the BCS superconductor $|g_{BCS}\rangle$ becomes:
$|\langle
g_F|g_{BCS}\rangle|=\prod_k\frac{1}{\sqrt{2}}\left(1+\frac{|\varepsilon_k-\varepsilon_F|}{\sqrt{(\varepsilon_k-\varepsilon_F)^2+\Delta(\delta
V)^2}}\right)^{1/2}$. We see that the BCS model at $T=0$ features the
Anderson orthogonality catastrophe, just as the Stoner-Hubbard
model.

\begin{figure}[ht]
\centering
\includegraphics[width=6.5cm,height=7.0cm,angle=-90]{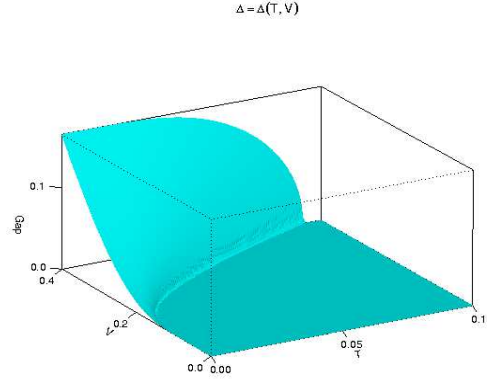}
\caption{(color online) The gap $\Delta=\Delta(T,V)$ as a function of the temperature $T$ and the coupling constant $V$. The plot is given in rescaled quantities $T\rightarrow k_BT/(\hbar\omega_D)$, $V\rightarrow D_FV$ and $\Delta \rightarrow \Delta/(\hbar\omega_D)$.} \label{BCS-Gap Figure}
\end{figure}

\begin{figure}[ht]
\centering
\includegraphics[width=5.5cm,height=7.0cm,angle=-90]{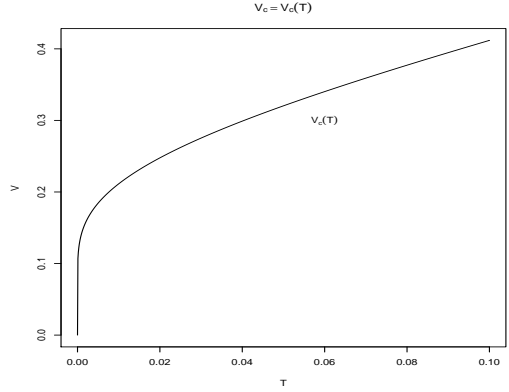}
\caption{(color online) The critical line $V_c=V_c(T)$ as a function of the temperature $T$. The plot is given in rescaled quantities $T\rightarrow k_BT/(\hbar\omega_D)$ and $V\rightarrow D_FV$.} \label{BCS_Uc Figure}
\end{figure}

\begin{figure}[ht]
\centering
\includegraphics[width=6.5cm,height=7.0cm,angle=-90]{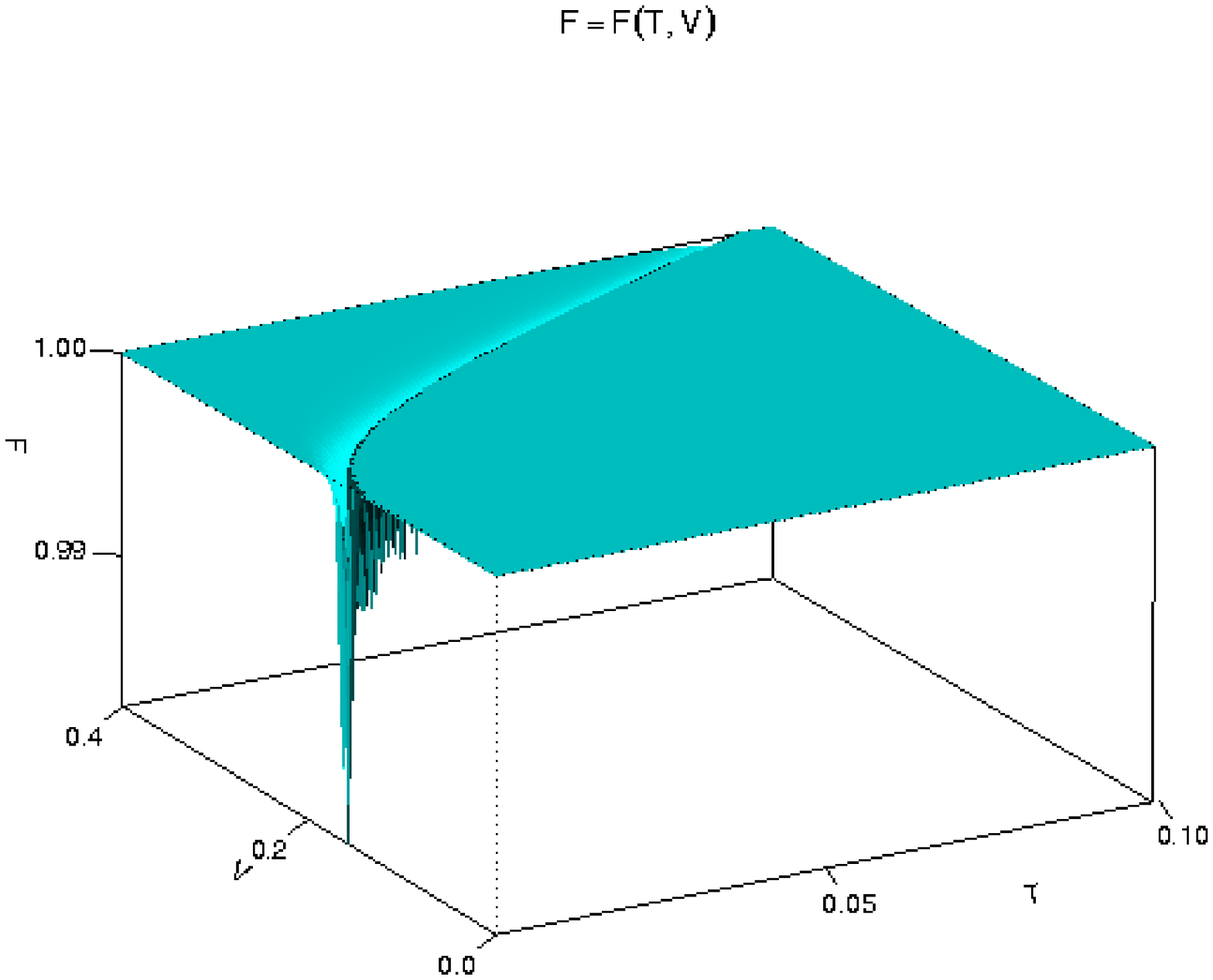}
\caption{(color online) The fidelity $F=F(T,V)$ as a function of the temperature $T$ and the coupling constant $V$. The plot is given in rescaled quantities $T\rightarrow k_BT/(\hbar\omega_D)$ and $V\rightarrow D_FV$, with $\delta T =0, \delta V = 10^{-3}$.} \label{BCS-Fidelity Figure}
\end{figure}

As in the previous case, in obtaining the numerical results for the
fidelity, we used the subroutine hybrd.f from MINPACK
\cite{programme}. Again, all the
numerical results are given in rescaled quantities $T\rightarrow
k_BT/(\hbar\omega_D)$, $V\rightarrow D_FV$ and $\Delta \rightarrow
\Delta/(\hbar\omega_D)$. The result for the gap is given on Fig. \ref{BCS-Gap Figure}. The line of the phase transition is clearly
marked as the line along which the gap becomes non-trivial, and is
presented on Fig. \ref{BCS_Uc Figure}. Finally, the fidelity, with
$\delta T =0, \delta V = 10^{-3}$, is plotted on Fig.
\ref{BCS-Fidelity Figure}. We varied the parameter differences
$\delta T$ and $\delta V$, and all the results obtained show
the same qualitative picture -- the fidelity exhibits a sudden drop
from $F=1$ precisely along the line of the phase transition.


As already discussed, in the case of mutually commuting Hamiltonians, the fidelity reduces to the quantity
\begin{equation}
\label{C_definition}
    C(\hat{\rho}_a, \hat{\rho}_b) = \frac{Z(\hat{\bar{H}})}{\sqrt{Z(\hat{\bar{H}}+\frac{\Delta\hat{H}}{2})Z(\hat{\bar{H}}-\frac{\Delta\hat{H}}{2})}}
\end{equation}
which, through relation (\ref{commuting_fidelity}), establishes the
connection between the singular behavior of the fidelity and the
corresponding susceptibility (or the heat capacity, etc.). In the
non-commuting case, the same relation between $C(\hat{\rho}_a,
\hat{\rho}_b)$ and $\chi$ is still valid (see Appendix $4$), yet the fidelity is in
general {\it not} identically equal to $C$.

\begin{figure}[ht]
\centering
\includegraphics[width=6.5cm,height=7.0cm,angle=-90]{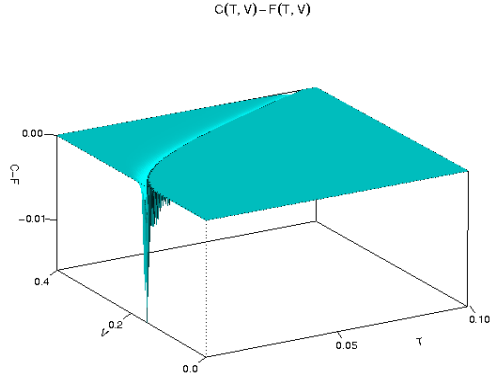}
\caption{(color online) The difference $C(T,V)-F(T,V)$ as a function of the temperature $T$ and the coupling constant $V$. The plot is given in rescaled quantities $T\rightarrow k_BT/(\hbar\omega_D)$ and $V\rightarrow D_FV$, with $\delta T =0, \delta V = 10^{-3}$.} \label{BCS-CF Figure}
\end{figure}

If we approach a line of the phase transition along a curve
$q=q(\alpha)$, $\alpha \in \mathbb{R}$, we find: $F=C(1+\mathcal{F}d\alpha^2)$, where $d\alpha$ defines the
difference $\delta q$. For reasons of simplicity, we omit here the
explicit, lengthy expression for $\mathcal{F}$. Although relatively
complicated and difficult to study directly, it is evident that,
apart from the Hamiltonian's eigenvalues, it is also explicitly
given by the rate of change, with respect to $\alpha$, of the
Hamiltonian's eigenbasis. This is also evident from the fact that
for the case of mutually commuting Hamiltonians, when the eigenbasis is common, $C\equiv F$ and
thus $\mathcal{F}\equiv 0$. In the commuting case, the change of
state is given by the change of the Hamiltonian's eigenvalues only,
while in the non-commuting case, the overall change of state is
given by the change of {\it both} the eigenvalues and the
eigenvectors. Thus, the singularity of $\mathcal{F}$ can on its own
mark the drastic change in the structure of the system's eigenbasis,
thus bringing about the finite difference between $C$ and $F$. This
is indeed the case for BCS superconductors, where the difference
$C-F$ becomes non-trivial precisely along the line of the phase
transition, where $C<F$, see Fig. \ref{BCS-CF Figure}. We see that,
as intuitively expected, the state of a system exhibits an abrupt
change in its structure along the line of the phase transition both
in terms of its eigenvalues, as well as in terms of its eigenstates
(for the structural analysis of the system's eigenstates given by a
parametrized Hamiltonian $\hat{H}(q)$, see for example \cite{cejnar}
and \cite{geisel}).


Another way to quantify the structural change of the eigenvectors is
through the Uhlmann connection and the corresponding mixed-state geometric phase
\cite{uhlmann}, the mixed-state generalization of the Berry connection and phase. 
The fidelity $F(\hat{\rho}_a, \hat{\rho}_b) = \mbox{Tr} \sqrt{\sqrt{\hat{\rho}_a}\hat{\rho}_b\sqrt{\hat{\rho}_a}}$
can also be expressed as
$F(\hat{\rho}_a, \hat{\rho}_b)=\mbox{Tr}\left[|\sqrt{\hat{\rho}_a}\sqrt{\hat{\rho}_b}|\right]$,
where $|\hat{A}|=(\hat{A}\hat{A}^\dag)^{1/2}$ represents the modulus of the operator $\hat{A}$ (see \cite{sjoqvist}). The Uhlmann parallel transport condition (i.e. the connection) is given by the unitary operator $\hat{U}_{ab}$, such that
$F(\hat{\rho}_a, \hat{\rho}_b)=\mbox{Tr}\left[\sqrt{\hat{\rho}_a}\sqrt{\hat{\rho}_b}\hat{U}_{ab}\right]$ (see equation ($9$) from \cite{sjoqvist}). In other words, the connection operator defining the parallel transport is the inverse of the unitary $\hat{V}_{ab}$, given by the polar decomposition \cite{nielsen} $\sqrt{\hat{\rho}_a}\sqrt{\hat{\rho}_b}=|\sqrt{\hat{\rho}_a}\sqrt{\hat{\rho}_b}|\hat{V}_{ab}$, i.e. $\hat{U}_{ab}=\hat{V}_{ab}^\dag$. Note that the parallel transport condition (the connection), given by $\hat{U}_{ab}$, induces both the local Uhlmann curvature two-form, as well as the global mixed-state geometric phase (see equations ($11$) and ($12$) from \cite{sjoqvist}). Let us define
\begin{equation}
    H(\hat{\rho}_a, \hat{\rho}_b) = \mbox{Tr}\left[\sqrt{\hat{\rho}_a}\sqrt{\hat{\rho}_b}\right].
\end{equation}
Obviously, in the case of mutually commuting Hamiltonians $H=F(=C)$
and the Uhlmann connection is trivial, $\hat{U}_{ab}=\hat{I}$. In
the case of the BCS model, we have evaluated the quantity $H-F$, and
the result is qualitatively the same as for $C-F$. In other words, along the line of the phase transition we have found the strict inequality
$H-F=\mbox{Tr}\left[|\sqrt{\hat{\rho}_a}\sqrt{\hat{\rho}_b}|(\hat{U}_{ab}-\hat{I})\right]<\mbox{const}<0$.
From the Cauchy-Schwarz inequality it follows then that
$\hat{U}_{ab}-\hat{I}\neq 0$, or $\hat{U}_{ab}\neq \hat{I}$ -- a clearly abrupt change of the connection operator $\hat{U}_{ab}$ occours and it becomes non-trivial in the vicinity of the line of the
phase transition. Since the two parameter points $q_a$ and $q_b$ are taken to be close to each other, such a behavior implies the non-analyticity of the local Uhlmann curvature form along the line of the phase transition, which in turn results in generally non-trivial global Uhlmann mixed-state geometric phase (see for example the discussions in \cite{zanardi-differential} and \cite{zanardi-scaling}). This can be seen as a mixed thermal state
generalization of recent results \cite{qpt-berry} on the relation
between QPTs and Berry geometric phases. Further, we have that along
the line of the phase transition we have $C<H<F$, while $C\simeq
H\simeq F\simeq 1$ otherwise.

\section{Conclusion}

In this paper we analyzed the fidelity approach to both zero
temperature (quantum) as well as finite temperature phase
transitions. It is based on the notion of quantum state
distinguishability, applied to the case of macroscopic many-body
systems whose states determine the global order of the system and
its phase. We focused on the two particular cases of the
Stoner-Hubbard model for itinerant electron magnetism and the BCS
theory of superconductivity, as representatives of two distinct
classes of physical systems: those defined by a set of mutually
commuting Hamiltonians and those defined by mutually {\em
non}-commuting Hamiltonians, with respect to a given parameter
space. We found that in both cases the fidelity can mark the regions
of PTs by its sudden drop in the vicinity of the transition line. We
discussed in detail the general case of mutually commuting
Hamiltonians, where in the case of symmetry-breaking LGW type of
transitions the non-analyticity of fidelity is a direct consequence
of the non-analyticity of the corresponding ``generalized''
susceptibility. The case of mutually non-commuting Hamiltonians is
more complex as there the structure of the Hamiltonian's
eigenvectors directly affects the system's thermal state thus
introducing a novel feature relevant for the macroscopic phase
distinguishability. On the example of the BCS superconductivity we
showed that the phase transition is accompanied by the abrupt change
of the Hamiltonian's eigenvectors, quantified by the differences
$C-F$ and $H-F$. The second one is of particular interest as it
determines the Uhlmann connection and the associated mixed-state geometric phase, which
generically becomes non-trivial in the vicinity of the phase transition only -- a
mixed state generalization of the emergence of non-trivial Berry
geometric phase in the vicinity of criticality for QPTs.

There are two main extensions of this work. First, the study of more
general phase transitions that fall outside the standard
symmetry-breaking LGW paradigm. Second, the general study of the
structural analysis of the system's eigenstates in the case of
mutually non-commuting Hamiltonians, given within the framework of
the mixed state fidelity and the Uhlmann geometric phase.

\section{Acknowledgments}

The authors thank EU FEDER and FCT projects QuantLog POCI/MAT/55796/2004 and QSec PTDC/EIA/67661/2006.
NP thanks the support from EU FEDER and FCT grant SFRH/BPD/31807/2006. 
VRV thanks the support of projects SFA-2-91, POCI/FIS/58746/2004 and FCT PTDC/FIS/70843/2006 and also useful discussions with Miguel A. N. Ara\'{u}jo
and Pedro D. Sacramento.

\section{Appendix $1$}

In this Appendix we prove various technical statements relevant for the analysis of the ground state fidelity in the Stoner-Hubbard model.

\begin{itemize}

\item {\bf Proving the relation $\varepsilon_F=\tilde{\varepsilon}_{k_{F\ua}}=\tilde{\varepsilon}_{k_{F\da}}$:}

Using $N_{\sigma}=\int_0^{N_\sigma} dN_\sigma$, the ground state energy can be written as:
\begin{eqnarray}
    E_g = \int_0^{N_\ua} E_{k\ua}dN_\ua + \int_0^{N_\da} E_{k\da}dN_\da - U\frac{N_\ua N_\da}{V} = \int_0^{N_\ua} \varepsilon_{k}dN_\ua + \int_0^{N_\da} \varepsilon_{k}dN_\da + U\frac{N_\ua N_\da}{V}.
\end{eqnarray}
Since it is the local minima, for given $N=N_\ua + N_\da$ we have that:
\begin{eqnarray}
    \frac{\partial (E_g - \mu N)}{\partial N_\ua} = \varepsilon_{k_F\ua} - \mu + U\frac{N_\da}{V} = 0, \nonumber \\
    \frac{\partial (E_g - \mu N)}{\partial N_\da} = \varepsilon_{k_F\da} - \mu + U\frac{N_\ua}{V} = 0,
\end{eqnarray}
and we immediately get $E_{k_{F\ua}} =
\varepsilon_{k_F\ua} + Un_\da = \varepsilon_{k_F\da} + Un_\ua =
E_{k_{F\da}}=\varepsilon_F$, with $\mu=\varepsilon_F$, for $T=0$.

\item {\bf Deriving $U_c$ from Stoner criterion:}

The density of states is defined as
$D(\varepsilon)=\frac{1}{V}\frac{dN_\sigma}{d\varepsilon}$ in the paramagnetic phase, where $\varepsilon$ is the
one-particle energy and $N_\sigma(\varepsilon)$ is the number of states for each spin direction
whose energy is smaller or equal than $\varepsilon$. For each spin projection, we have $N_\sigma/V=k_F^3/(6\pi^2)$,
while from $\varepsilon(k)=\frac{\hbar^2}{2m}k^2$ we have that
$k=(\frac{2m}{\hbar^2}\varepsilon)^{1/2}$. Thus,
$\frac{N_\sigma(\varepsilon)}{V}=\frac{1}{6\pi^2}(\frac{2m}{\hbar^2}\varepsilon)^{3/2}$
and:
\begin{equation}
\label{density}
D(\varepsilon)=\frac{1}{V}\frac{dN_\sigma(\varepsilon)}{d\varepsilon}=\frac{3}{2}\frac{1}{6\pi^2}\left(\frac{2m}{\hbar^2}\right)^{3/2}\sqrt{\varepsilon}=\frac{3}{2\varepsilon}\frac{N_{\sigma}}{V}=\frac{3}{4\varepsilon}\frac{N}{V}.
\end{equation}

The difference between the up and down single particle energies $E_{k\ua}$ and $E_{k\da}$ defines the self consistent magnetic field $\Delta=2U\frac{M(\Delta)}{V}$. The phase transition occurs when $\Delta$ becomes non-zero, leading to the equation $1=\frac{2U_c}{V}\frac{\partial M}{\partial \Delta}|_{\Delta=0}$, which also gives the vanishing of the susceptibility denominator in the random phase approximation (Stoner enhancement factor) \cite{stoner}. One has then $U_cD_F=1$, and therefore we find $U_c=\frac{1}{D_F}=\frac{4}{3}\varepsilon_F\frac{V}{N}$.

\item {\bf Deriving $U_c$ from (\ref{SH-eq1}) and (\ref{SH-eq2}):}

Using $x=k_{F\ua}$, $y=k_{F\da}$, equation (\ref{SH-eq1}) can be
rewritten as $f(x,y)\equiv x^3+y^3-a=0$, with $a =
6\pi^2\frac{N}{V}\equiv 2k_F^3$. Also, the quadratic term from
equation (\ref{SH-eq2}) that is for $U>U_c$ responsible for
non-trivial magnetic solutions reads as $g(x,y;U)\equiv (x+y) -
\frac{U}{\alpha}(x^2+xy+y^2) = 0$, with $\alpha =
3\pi^2\frac{\hbar^2}{m}\equiv
\frac{3}{2}k_FU_c$.

First, we show that for
$U>U_c$ (see equation
(\ref{U_c})), curves $f$ and $g$ have intersection in two points,
symmetric with respect to the $y=x$ line, while for $U=U_c$ they touch
precisely in the point $(x,y)=(k_F,k_F)$. In the coordinate system
$(\bar{x},\bar{y})$, rotated by the angle $\varphi = \pi/4$ and
translated by the vector (A,0) from the system (x,y), the curve
$g(\bar{x},\bar{y};U)=0$ becomes \footnote{Alternatively, we could
first translate the coordinate system by the vector
$(\sqrt{2}A,\sqrt{2}A)$ and then rotate it by the angle $\varphi =
\pi/4$ to obtain the same system $(\bar{x},\bar{y})$.}:
\begin{equation}
    \frac{\bar{x}^2}{A^2}+\frac{\bar{y}^2}{B^2}=1,
\end{equation}
with $A=\frac{\sqrt{2}}{3}\frac{k_F}{U}$ and
$B=\sqrt{\frac{2}{3}}\frac{k_F}{U}$ -- a real ellipse with the main
axes $A=A(U)$ and $B=B(U)$. As the parameter $U$ increases from $0$
(when $A,B\rightarrow\infty)$, the main axes $A$ and $B$ decrease
and eventually the ellipse $g$ touches the curve $f$ in the point
$(\bar{x},\bar{y})=(2A,0)$, which in the original coordinate system
correspond to the point $(x,y)=(k_F,k_F)$. In other words,
$2A=\sqrt{2}k_F$ and $U=U_c$. Further increase of $U$ beyond $U_c$
results in even smaller ellipses that intersect the curve $f$ in two
symmetric points (note that the curve $f$ is itself symmetric along
the $y=x$ line).

To show that for $U>U_c$ the ``balanced'' solutions of
(\ref{SH-eq1}) and (\ref{SH-eq2}) are indeed the physical ones, we
calculate the total ground state energy $E_g$ of the system:
\begin{equation}
E_g=V\int_{E_{0\ua}}^{E_{F\ua}}\tilde{\varepsilon}_{k\ua}D(E_{k\ua})dE_{k\ua} + V\int_{E_{0\da}}^{E_{F\da}}E_{k\da}D(E_{k\da})dE_{k\da} - VUn_{\ua}n_{\da}.
\end{equation}
Upon evaluating the above integrals, using $M=(N_\ua-N_\da)/2$, the total ground state energy reads:
\begin{equation}
    E_g=\frac{V}{20\pi^2}\frac{\hbar^2}{m}\left(3\pi^2\frac{N}{V}\right)^{5/3}\left[\left(1+\frac{2M}{N}\right)^{5/3} + \left(1-\frac{2M}{N}\right)^{5/3}\right] + \frac{UN^2}{4V}\left[1-\left(\frac{2M}{N}\right)^2\right].
\end{equation}
From the above expression, we see that for every value of the
coupling constant $U$, $\frac{\partial E_g}{\partial M}(U,M=0)=0$,
while $\frac{\partial^2 E_g}{\partial M^2}(U,M=0)$ changes the sign
from positive to negative precisely in $U=U_c$. Thus, we indeed have
that for $U>U_c$ the ``balanced'' magnetic solutions with $M\neq 0$
have lower energy.

\item {\bf Deriving the maximum fidelity, i.e. the maximum derivative $d k_F/ d U$ for $U=U_c$, from (\ref{SH-eq1}) and (\ref{SH-eq2}):}

From $df=0$ and $dg=0$, we get ($\frac{\partial f}{\partial x} = f_x$, $\frac{d x}{d U} = \dot{x}$, etc.):
\begin{eqnarray}
    f_x\dot{x}+f_y\dot{y} & = & 0, \nonumber \\
    g_x\dot{x}+g_y\dot{y} & = & -g_u.
\end{eqnarray}
Solving it for $\dot{x}$ and $\dot{y}$, we obtain:
\begin{equation}
\left[\begin{array}{cc}
    \dot{x} \\ \dot{y}
\end{array}\right] = \frac{g_u}{f_xg_y-f_yg_x}
\left[\begin{array}{cc}
    f_y \\ -f_x
\end{array}\right].
\end{equation}
Finally:
\begin{equation}
\left[\begin{array}{cc}
    \dot{x} \\ \dot{y}
\end{array}\right] = \frac{3}{4}\frac{k_f}{D_FV^2}\frac{x+y}{x-y}\frac{1}{xy}
\left[\begin{array}{cc}
    y^2 \\ -x^2
\end{array}\right].
\end{equation}

Thus, in the limit $U\rightarrow U_{c+}$ (note that $g(x,y;U)=0$ is
valid for $U\geq U_c$), we have that $x\rightarrow k_{F+}$,
$y\rightarrow k_{F-}$ and $\dot{x}\rightarrow +\infty$,
$\dot{y}\rightarrow -\infty$ (or vice versa, for the $x<y$ solution
branch). In other words, $\frac{d V(k_{F\da})}{d U} =
4\pi(k_{F\da})^2\frac{d(k_{F\da})}{dU} = 4\pi x^2\dot{x} \rightarrow
+\infty$ and $\frac{d V(k_{F\ua})}{d U} \rightarrow -\infty$, or
vice versa.

\end{itemize}

\section{Appendix $2$}

Below, the basic identities of the $\mbox{su}(2)$ algebra of the
electron spin operators leading to equation
(\ref{SH_fidelity-final}) are proven. We discuss operators having a
fixed momentum and drop the index $k$ for convenience.

\begin{itemize}

\item {\bf $\hat{\vec{S}}$-algebra identities:}

In general, operators defined by $\hat{S}^a = \hat{c}^\dag_\alpha
S^a_{\alpha\beta}\hat{c}_\beta$, where $\hat{c}^\dag_\alpha$ and
$\hat{c}_\beta$ are either bosonic or fermionic operators and
$S^a_{\alpha\beta}$ is a matrix representation of a spin $S$
algebra, i.e. satisfying
$[S^a,S^b]_{\alpha\beta}=i\varepsilon^{abc}S^c_{\alpha\beta}$,
define a $\mbox{su}(2)$ algebra. This is the case of the electron
spin operators defined by
$\hat{\vec{S}}=\hat{\Psi}^\dag\frac{\vec{\sigma}}{2}\hat{\Psi}$,
where $\hat{\Psi}^\dag = [\hat{c}_{\ua}^\dag  \ \hat{c}_{\da}^\dag
]$ and $\vec{\sigma}$ is the vector of Pauli matrices, and given by
$\hat{S}^z=\frac{1}{2}(\hat{n}_{\ua}-\hat{n}_{\da})$,
$\hat{S}^+=\hat{c}_{\ua}^\dag\hat{c}_{\da}$ and
$\hat{S}^-=\hat{c}_{\da}^\dag\hat{c}_{\ua}$ (with $\hat{S}^\pm =
\hat{S}^x\pm i\hat{S}^y$).

Using the anti-commutation relations for the one-electron modes
$\hat{c}^\dag_{\sigma}$, one can easily obtain the following
commutation and anti-commutation identities:
\begin{eqnarray}\label{s-algebra-id1}
     [\hat{S}^z,\hat{S}^\pm] = \pm\hat{S}^\pm, & & [\hat{S}^+,\hat{S}^-] = 2\hat{S}^z, \\
    \{\hat{S}^z,\hat{S}^\pm\} = 0, & & \{\hat{S}^+,\hat{S}^-\} = \hat{I}_s. \nonumber
\end{eqnarray}
with $\hat{I}_s \equiv
\hat{n}_{\ua}+\hat{n}_{\da}-2\hat{n}_{\ua}\hat{n}_{\da} = (2-\hat{n})\hat{n}$, $\hat{n}=\hat{n}_{\ua}+\hat{n}_{\da}$, the projector onto the subspace of the single-occupied states, being an
identity operator for this algebra, since $\hat{I}_s^2 = \hat{I}_s$,
$\hat{I}_s\hat{\vec{S}} =\hat{\vec{S}} \hat{I}_s=\hat{\vec{S}}$ and
$(\hat{\vec{S}})^2=\frac{3}{4}\hat{I}_s$. One of the immediate
consequences of the above commutation relations is that
$\{\hat{S}^x,\hat{S}^y,\hat{S}^z\}$ form a $\mbox{su}(2)$ algebra.

Further, we can obtain the traces: $\mbox{Tr}[\hat{I}] = 4$, $\mbox{Tr}[\hat{I}_s]=2$ and $\mbox{Tr}[\hat{\vec{S}}] = \vec{0}$.

\item {\bf Evaluating $e^{\vec{h}\hat{\vec{S}}}$ and} $\mbox{Tr}[e^{\vec{h}\hat{\vec{S}}}]${\bf :}

\begin{equation}
    e^{\vec{h}\hat{\vec{S}}} = \sum_{n=0}^{+\infty} \frac{1}{n!}(\vec{h}\hat{\vec{S}})^n = \hat{I} + \sum_{n=1}^{+\infty}\frac{1}{n!}(\vec{h}\hat{\vec{S}})^n.
\end{equation}
For $(\vec{h}\hat{\vec{S}})^2$, we have:
\begin{eqnarray}
\label{vec_s_prod1}
    (\vec{h}\hat{\vec{S}})^2 & = & [\frac{1}{2}(h^+\hat{S}^- + h^-\hat{S}^+)+h_z\hat{S}^z]^2 \\
   & = & \frac{1}{4}(h^+)^2(\hat{S}^-)^2 + \frac{1}{4}(h^-)^2(\hat{S}^+)^2 + h_z^2(\hat{S}^z)^2 + \frac{1}{4}(h^+h^-)\{\hat{S}^-,\hat{S}^+\} + \frac{1}{2}(h_zh^+)\{\hat{S}^z,\hat{S}^-\} + \frac{1}{2}(h_zh^-)\{\hat{S}^z,\hat{S}^+\}. \nonumber
\end{eqnarray}
Using the $\hat{\vec{S}}$-algebra identities (\ref{s-algebra-id1}), we get:
\begin{equation}
\label{vec_s_prod2}
    (\vec{h}\hat{\vec{S}})^2 = \frac{1}{4}(h_z)^2\hat{I}_s + \frac{1}{4}(h^+h^-)\hat{I}_s = \frac{1}{4}|h|^2\hat{I}_s,
\end{equation}
where \footnote{Using $A^{\pm}=A_x\pm iA_y$ and $A^0=A_z$, we can
write $\vec{A}=A_x\vec{e}_x + A_y\vec{e}_y + A_z\vec{e}_z =
A^{+}\vec{e}_{+} + A^{-}\vec{e}_{-} + A^0\vec{e}_0$, where
$\vec{e}_{\pm} = \vec{e}_x \pm i\vec{e}_y$ and $\vec{e}_0=\vec{e}_z$
(and analogously for $\vec{B}$). The scalar product between two
vectors is then
$\vec{A}\cdot\vec{B}=A_xB_x+A_yB_y+A_zB_z=\frac{1}{2}(A^{+}B^{-}+A^{-}B^{+})+A^0B^0$.
Therefore, $A_k^2=A^{+}A^{-}+(A^0)^2$.} $|h|^2 = (h_z)^2 + h^+h^-
\equiv h^2$. Thus, we obtain:
\begin{eqnarray}
\label{one_exponent_s}
    e^{\vec{h}\hat{\vec{S}}} & = & \hat{I} + \sum_{n=1}^{+\infty}\frac{1}{(2n)!}\left(\frac{h}{2}\right)^{2n}\hat{I}_s + \sum_{n=0}^{+\infty}\frac{1}{(2n+1)!}\left(\frac{h}{2}\right)^{2n}\!\!(\vec{h}\hat{\vec{S}}) \nonumber \\ & = & (\hat{I} - \hat{I}_s) + \left[\sum_{n=0}^{+\infty}\frac{1}{(2n)!}\left(\frac{h}{2}\right)^{2n}\right]\hat{I}_s  + \left[\sum_{n=0}^{+\infty}\frac{1}{(2n+1)!}\left(\frac{h}{2}\right)^{2n+1}\right]\frac{2}{h}(\vec{h}\hat{\vec{S}}), \nonumber \\ e^{\vec{h}\hat{\vec{S}}} & = & (\hat{I} \! - \! \hat{I}_s) + \cosh\left(\frac{h}{2}\right)\hat{I}_s + 2\sinh\left(\frac{h}{2}\right)\frac{(\vec{h}\hat{\vec{S}})}{h}.
\end{eqnarray}

Using the trace formulas, we finally obtain:
\begin{equation}
\label{s_trace}
    \mbox{Tr}[e^{\vec{h}\hat{\vec{S}}}] = 2\left(1 + \cosh\frac{h}{2}\right).
\end{equation}

\item {\bf Evaluating $e^{\alpha\hat{n}}$ and} $\mbox{Tr}[e^{\alpha\hat{n}}]${\bf :}

Enlarging the $\mbox{su}(2)$ algebra of the $\hat{\vec{S}}$
operators to the operator $\hat{n}=\hat{n}_{\ua}+\hat{n}_{\da}$ it
is useful to know its algebraic properties and their relation to
the other operators.
We first evaluate $\hat{n}^k$. Since
$\hat{n}^2_{\sigma}=\hat{n}_{\sigma}$ and
$\hat{n}^2=(\hat{n}_\ua+\hat{n}_\da)^2=\hat{n}+2\hat{n}_\ua\hat{n}_\da$
we find that $\hat{n}^3=-2\hat{n}+3\hat{n}^2$ and in general
$\hat{n}^k = -a_k\hat{n}+b_k\hat{n}^2$, where $a_k$ and $b_k$
satisfy the recurrence equations $a_{k+1}=2b_k$ and
$b_{k+1}=3b_k-a_k$, leading to:
$\hat{n}^k=-(2^{k-1}-2)\hat{n}+(2^{k-1}-1)\hat{n}^2$. The
exponential $e^{\alpha\hat{n}}$ is given by:
\begin{eqnarray}
e^{\alpha\hat{n}} = \hat{I}-\left(\frac{e^{2\alpha}}{2}-2e^{\alpha}+\frac{3}{2}\right)\hat{n}+\left(\frac{e^{2\alpha}}{2}-e^{\alpha}+\frac{1}{2}\right)\hat{n}^2 = \hat{I}+u\hat{n}+v\hat{n}^2,
\end{eqnarray}
which can be easily verified since $\hat{n}=\hat{n}_{\ua}+\hat{n}_{\da}$ can only take the values $n=0,1,2$.

Using the obvious trace relations $\mbox{Tr}[\hat{n}]=4$ and $\mbox{Tr}[\hat{n}^2]=6$, we also get $\mbox{Tr}[e^{\alpha\hat{n}}]=4+4u+6v$, or:
\begin{equation}
    \mbox{Tr}[e^{\alpha\hat{n}}]=(e^{\alpha}+1)^2.
\end{equation}

\end{itemize}

The number operator commutes with the $\hat{\vec{S}}$ operators,
and one has $\hat{n}\hat{I}_s=\hat{I}_s\hat{n}=\hat{I}_s$ and
$\hat{n}\hat{\vec{S}}=\hat{\vec{S}}\hat{n}=\hat{\vec{S}}$. Finally,
one obtains:
\begin{equation}
    \mbox{Tr}\left[e^{(\alpha\hat{n}+\vec{h}\hat{\vec{S}})}\right] = \left(1+e^{\alpha+\frac{h}{2}}\right)\left(1+e^{\alpha-\frac{h}{2}}\right) = 2 \ e^\alpha\left( \cosh\alpha+\cosh\frac{h}{2}\right).
\end{equation}

\section{Appendix $3$}

In the following, we obtain the basic formulas for the
$\mbox{su}(2)$ algebra of the Nambu operators $\hat{\vec{T}}$ used
in evaluating equation (\ref{BCS-fidelity}) for the fidelity between
two states of a BCS superconductor in thermodynamical
equilibrium.

\begin{itemize}

\item {\bf $\hat{\vec{T}}$-algebra identities:}

We now consider the operators defined by $\hat{\vec{T}} =
\hat{\psi}^\dag \frac{\vec{\sigma}}{2} \hat{\psi}$, where $\hat{\psi}^\dag = [\hat{c}_{\ua}^\dag  \ \hat{c}_{\da} ]$ and
$\vec{\sigma}$ is the vector of Pauli matrices, and given by $\hat{T}^z=\frac{1}{2}(\hat{n}_\ua+\hat{n}_\da-1)$, $\hat{T}^+=\hat{c}^\dag_\ua\hat{c}^\dag_\da$ and $\hat{T}^-=\hat{c}_\da\hat{c}_\ua$ (with $\hat{T}^{\pm}=\hat{T}^x\pm i\hat{T}^y$). These operators $\hat{T}^a$ are of the form considered before, since the fermion anti-commutation relations are invariant under the electron-hole transformation, i.e. under the interchange $\hat{c} \leftrightarrow \hat{c}^\dag$. Similarly to the previous case of the $\hat{\vec{S}}$-algebra, we obtain the commutation and anti-commutation relations:
\begin{eqnarray}\label{t-algebra-id1}
     [\hat{T}^0,\hat{T}^\pm] = \pm\hat{T}^\pm, & & [\hat{T}^+,\hat{T}^-] = 2\hat{T}^0, \\
    \{\hat{T}^0,\hat{T}^\pm\} = 0, & & \{\hat{T}^+,\hat{T}^-\} = \hat{I}_t. \nonumber
\end{eqnarray}
with $\hat{I}_t \equiv 2\hat{n}_{\ua}\hat{n}_{\da}-(\hat{n}_{\ua}+\hat{n}_{\da})+1=(n-1)^2$, the projector onto the subspace of the empty and doubly occupied states, being an identity operator for this algebra, since $\hat{I}_t^2 = \hat{I}_t$, $\hat{I}_t\hat{\vec{T}} =\hat{\vec{T}} \hat{I}_t=\hat{\vec{T}}$ and $(\hat{\vec{T}})^2 = \frac{3}{4}\hat{I}_t$. One of the immediate consequences of the above commutation relations
is that $\{\hat{T}^0,\hat{T}^+,\hat{T}^-\}$ form a
$\mbox{su}(2)$ algebra. Further, simple algebra gives:
\begin{equation}\label{t-algebra-id2}
    \mbox{Tr}[\hat{\vec{T}}] = 0, \ \ \ \ \mbox{Tr}[\hat{I}_t]=2.
\end{equation}

Note that in the above definition of $\hat{\psi}^\dag$, the operators $\hat{c}_{\ua}^\dag$ and $\hat{c}_{\da}$ have the opposite momenta, $k$ and $-k$, respectively. Had the momenta been defined to be the same, the two sets of operators $\hat{\vec{S}}$ and $\hat{\vec{T}}$ would commute with each other, i.e. $[\hat{S}^a,\hat{T}^b]=0$. Also, one would have $\hat{I}_s+\hat{I}_t=\hat{I}$.

\item {\bf Evaluating $e^{\vec{a}\hat{\vec{T}}}$ and} $\mbox{Tr}[e^{\vec{a}\hat{\vec{T}}}]${\bf :}

\begin{equation}
    e^{\vec{a}\hat{\vec{T}}} = \sum_{n=0}^{+\infty} \frac{1}{n!}(\vec{a}\hat{\vec{T}})^n = \hat{I} + \sum_{n=1}^{+\infty}\frac{1}{n!}(\vec{a}\hat{\vec{T}})^n.
\end{equation}
Similarly to the previous case, for $(\vec{a}\hat{\vec{T}})^2$, we have:
\begin{equation}
\label{vec_t_prod1}
    (\vec{a}\hat{\vec{T}})^2 = [a_z\hat{T}^0 + \frac{1}{2}(a^+\hat{T}^- + a^-\hat{T}^+)]^2= \frac{1}{4}|a|^2\hat{I}_t,
\end{equation}
where $|a|^2 = (a_z)^2 + a^+a^- \equiv a^2$. Thus, we obtain ($\hat{Z}_t=\hat{I}-\hat{I}_t$):
\begin{eqnarray}
\label{one_exponent}
    e^{\vec{a}\hat{\vec{T}}} & = & \hat{I} +\sum_{n=1}^{+\infty}\frac{1}{(2n)!}\left(\frac{a}{2}\right)^{2n}\hat{I}_t + \sum_{n=0}^{+\infty}\frac{1}{(2n+1)!}\left(\frac{a}{2}\right)^{2n}(\vec{a}\hat{\vec{T}}) \nonumber \\ & = & (\hat{I} \! - \! \hat{I}_t) + \left[\sum_{n=0}^{+\infty}\frac{1}{(2n)!}\left(\frac{a}{2}\right)^{2n}\right]\hat{I}_t + \left[\sum_{n=0}^{+\infty}\frac{1}{(2n+1)!}\left(\frac{a}{2}\right)^{2n+1}\right]\frac{2}{a}(\vec{a}\hat{\vec{T}}), \nonumber \\ e^{\vec{a}\hat{\vec{T}}} \!\! & = & \!\! \hat{Z}_t + \cosh\left(\frac{a}{2}\right)\hat{I}_t + 2\sinh\left(\frac{a}{2}\right)\frac{(\vec{a}\hat{\vec{T}})}{a}.
\end{eqnarray}

Using the trace formulas (\ref{t-algebra-id2}), we finally obtain:
\begin{equation}
\label{trace}
    \mbox{Tr}[e^{\vec{a}\hat{\vec{T}}}] = 2\left(1 + \cosh\frac{a}{2}\right).
\end{equation}

\item {\bf Evaluating $e^{2\vec{c} \hat{\vec{T}}} = e^{\frac{\vec{a}}{2} \hat{\vec{T}}} e^{\vec{b} \hat{\vec{T}}} e^{\frac{\vec{a}}{2} \hat{\vec{T}}}$ and} $\mbox{Tr}[e^{\vec{c} \hat{\vec{T}}}]${\bf :}

From the above formula (\ref{trace}) and the known trigonometric
relation $(\cosh\frac{\alpha}{2})^2=\frac{1}{2}(1+\cosh\alpha)$, we
have that $\mbox{Tr}[e^{\vec{c} \hat{\vec{T}}}]=2\left(1 +
\cosh\frac{c}{2}\right)=2\left(1+\sqrt{\frac{1}{2}(1+\cosh c)}\right)$.
Thus, by evaluating $e^{2\vec{c} \hat{\vec{T}}}$ we obtain the result
for $\cosh c$ and therefore $\mbox{Tr}[e^{\vec{c}
\hat{\vec{T}}}]$. First, we obtain the general expression for
$e^{\vec{a}\hat{\vec{T}}}e^{\vec{b}\hat{\vec{T}}}$. Using the result
(\ref{one_exponent}) for $e^{\vec{a}\hat{\vec{T}}}$ (and analogously
for $e^{\vec{b}\hat{\vec{T}}}$), we get:
\begin{eqnarray}
    e^{\vec{a}\hat{\vec{T}}}e^{\vec{b}\hat{\vec{T}}} \!\!= \!\!\hat{Z}_t \! + \! \left(\cosh\frac{a}{2}\cosh\frac{b}{2}\right)\hat{I}_t \! + \!  2\left[\left(\cosh\frac{a}{2}\sinh\frac{b}{2}\right)\frac{\vec{b}}{b} \! + \!  \left(\sinh\frac{a}{2}\cosh\frac{b}{2}\frac{\vec{a}}{a}\right)\right]\hat{\vec{T}} \! + \! 4\!\left(\!\sinh\frac{a}{2}\sinh\frac{b}{2}\right)\frac{\vec{a}\hat{\vec{T}}}{a}\frac{\vec{b}\hat{\vec{T}}}{b}.
\end{eqnarray}
In deriving the above expression, we have used the above identities
(\ref{t-algebra-id2}). As in (\ref{vec_t_prod1}), using
(\ref{t-algebra-id1}) we obtain:
\begin{equation}
    (\vec{a}\hat{\vec{T}})(\vec{b}\hat{\vec{T}}) = \frac{1}{4}\left[(\vec{a}\vec{b})\hat{I}_t + 2i(\vec{a}\times\vec{b})\hat{\vec{T}} \right].
\end{equation}
Therefore, we have:
\begin{eqnarray}
    e^{\vec{a}\hat{\vec{T}}}e^{\vec{b}\hat{\vec{T}}} = \hat{Z}_t & + & \left[\left(\cosh\frac{a}{2}\cosh\frac{b}{2}\right) + \left(\sinh\frac{a}{2}\sinh\frac{b}{2}\right)\frac{(\vec{a}\vec{b})}{ab}\right]\hat{I}_t \\ & + &  2\left[\left(\sinh\frac{a}{2}\cosh\frac{b}{2}\right)\frac{\vec{a}}{a} + \left(\cosh\frac{a}{2}\sinh\frac{b}{2}\right)\frac{\vec{b}}{b} + i\frac{(\vec{a}\times\vec{b})}{ab} \right]\hat{\vec{T}}. \nonumber
\end{eqnarray}
Applying the above result twice, we finally obtain the expression
for $e^{2\vec{c} \hat{\vec{T}}}$:
\begin{eqnarray}
    e^{2\vec{c} \hat{\vec{T}}} & = & e^{\frac{\vec{a}}{2} \hat{\vec{T}}} e^{\vec{b} \hat{\vec{T}}} e^{\frac{\vec{a}}{2} \hat{\vec{T}}}  =   \hat{Z}_t +  \left[\left(\cosh\frac{a}{2}\cosh\frac{b}{2}\right) + \left(\sinh\frac{a}{2}\sinh\frac{b}{2}\right)\frac{(\vec{a}\vec{b})}{ab}\right]\hat{I}_t \\ & + & \left[\left(\sinh\frac{a}{2}\cosh\frac{b}{2}\right)\frac{\vec{a}}{a} + 2\left(\sinh\frac{b}{2}\right)^2\frac{\vec{b}}{b} +\left(\cosh\frac{a}{2}-1\right)\left(\sinh\frac{b}{2}\right)\frac{(\vec{a}\vec{b})\vec{a}}{a^2b}\right]\hat{\vec{T}}. \nonumber
\end{eqnarray}
Comparing the above result with the expression (\ref{one_exponent}),
we eventually end up with the expression for $\cosh c$:
\begin{equation}
    \cosh c = \left(\cosh\frac{a}{2}\cosh\frac{b}{2}\right) + \left(\sinh\frac{a}{2}\sinh\frac{b}{2}\right)\frac{(\vec{a}\vec{b})}{ab}
\end{equation}
which reduces to (\ref{cosh-explicit}) for particular values of $\vec{a}_k=\vec{\tilde{h}}_k(q_a)$ and $\vec{b}_k=\vec{\tilde{h}}_k(q_b)$.
\end{itemize}

\section{Appendix $4$}

In this Appendix we prove that in the case of mutually
non-commuting Hamiltonians a relation analogous to
(\ref{commuting_fidelity}) holds between $C$, given by equation
(\ref{C_definition}), and the susceptibility $\chi$. As in the
commuting case, for simplicity, we consider a Hamiltonian
$\hat{H}=\hat{H}_0 - h\hat{S}$ with the symmetry-breaking term
$\hat{S}$, and $h=h(q)$. Note that in this case, the two terms in
the Hamiltonian {\em do not commute} with each other,
$[\hat{H}_0,\hat{S}]\neq 0$. Thus, we have the following imaginary time Dyson expansion around the point $h=0$
\footnote{We consider the particular case of the $h=0$ expansion,
but all the results obtained can be easily generalized for the
$h\neq 0$ case.}:
\begin{equation}
 \label{A4_1}
e^{-\beta(\hat{H}_0 - h\hat{S})} \simeq \left\{ e^{-\beta\hat{H}_0} + h\int_0^\beta d\tau e^{-\beta\hat{H}_0}\hat{S}(\tau) + h^2\int_0^\beta d\tau \int_0^\tau d\tau_1e^{-\beta\hat{H}_0}\hat{S}(\tau)\hat{S}(\tau_1) \right\},
\end{equation}
with $\hat{S}(\tau)=e^{\tau\hat{H}_0}\hat{S}e^{-\tau\hat{H}_0}$.
From the above equation, we obtain the expressions for the magnetization $M=\langle\hat{S}\rangle$ and the susceptibility $\chi=\frac{\partial M}{\partial h}$ given by derivatives of the partition function $Z$. First, the magnetization can be expressed as (using the commutativity between the partial derivative and the trace, $\frac{\partial}{\partial h}\mbox{Tr} [\cdot]=\mbox{Tr}\frac{\partial}{\partial h} [\cdot]$): 
\begin{eqnarray}
M = \frac{1}{\beta}\frac{\partial\ln Z}{\partial h}=\frac{1}{\beta}\frac{1}{Z}\frac{\partial Z}{\partial h}=\frac{1}{\beta}\frac{1}{Z}\frac{\partial}{\partial h}\mbox{Tr} [e^{-\beta\hat{H}}]=\frac{1}{\beta}\frac{1}{Z}\int_0^\beta d\tau \mbox{Tr}[e^{-\beta\hat{H}_0}e^{\tau\hat{H}_0}\hat{S}e^{-\tau\hat{H}_0}]=\frac{1}{\beta}\int_0^\beta d\tau \mbox{Tr}[\frac{e^{-\beta\hat{H}_0}}{Z}\hat{S}]=\langle\hat{S}\rangle.
\end{eqnarray}
The susceptibility is then:
\begin{equation}
 \chi = \frac{\partial M}{\partial h} = \frac{\partial}{\partial h}\left(\frac{1}{\beta}\frac{1}{Z}\frac{\partial Z}{\partial h}\right) = \frac{1}{\beta}\frac{1}{Z}\frac{\partial^2 Z}{\partial h^2} - \frac{1}{\beta}\frac{1}{Z^2}\left(\frac{\partial Z}{\partial h}\right)^2.
\end{equation}
The second term is obviously equal to $\beta M^2=\beta\langle\hat{S}\rangle^2$, while the first term can be transformed as follows \footnote{In order to avoid a possible confusion, here we explicitly denote that the derivatives are taken for $h=0$.}:
\begin{eqnarray}
\!\! \frac{1}{\beta}\frac{1}{Z}\left(\frac{\partial^2 Z}{\partial h^2}\right)_{0} \!\! = \frac{1}{\beta}\frac{1}{Z}\left(\frac{\partial}{\partial h}\int_0^\beta \! d\tau \mbox{Tr}[e^{-\beta\hat{H}}\hat{S}]\right)_{0} \!\! = \frac{1}{Z} \mbox{Tr}[\left(\frac{\partial e^{-\beta\hat{H}}}{\partial h}\right)_{0}\hat{S}] = \frac{1}{Z}\int_0^\beta \!\! d\tau \mbox{Tr}[e^{-\beta\hat{H}_0}\hat{S}(\tau)\hat{S}] =  \!\! \int_0^\beta \!\!  d\tau\langle\hat{S}(\tau)\hat{S}\rangle.
\end{eqnarray}
Thus, the susceptibility is given by $\chi=\int_0^\beta d\tau [\langle\hat{S}(\tau)\hat{S}\rangle - \langle\hat{S}\rangle^2]$. From this, the Taylor expansion for $Z$, 
\begin{equation}
Z \simeq Z_0\left\{ 1 + \beta M h + \frac{1}{2}\beta^2 M^2 h^2 + \frac{1}{2}\beta\chi h^2 \right\},
\end{equation}
is identical to the one obtained for the case of mutually commuting Hamiltonians, and therefore a relation analog to equation (\ref{commuting_fidelity}) holds
between $C$ and the susceptibility $\chi$.

\end{document}